\documentclass[preprint,12pt]{elsarticle}

\usepackage[a4paper, left=1in, right=1in, top=1in, bottom=1in]{geometry}
\biboptions{sort&compress}

\usepackage{url}
\usepackage{amsmath}
\usepackage{amssymb}
\usepackage{graphicx}
\usepackage{textcomp}
\usepackage{hyperref}
 \usepackage{booktabs}
\usepackage{multicol}
\usepackage{multirow}
\usepackage[ruled,linesnumbered]{algorithm2e}
\usepackage{algpseudocode}
\usepackage{xcolor,colortbl}
\usepackage{adjustbox}
\usepackage{ragged2e}
\usepackage{array}
\usepackage{mathtools}
\usepackage{rotating}
\usepackage[most]{tcolorbox}
\usepackage{subfigure}
\usepackage{listings}
\usepackage{enumitem}
\usepackage{colortbl} 
\usepackage{threeparttable}
\usepackage{arydshln} 

\definecolor{mycolor_box}{HTML}{F5FFFA} 
\definecolor{mycolor_title}{HTML}{FFEBCD} 

\AtBeginDocument{%
  }

\begin{document}

\begin{frontmatter}




\title{
A Metamorphic Testing Perspective on Knowledge Distillation for Language Models of Code: Does the Student Deeply Mimic the Teacher?}


\author[1]{Md. Abdul Awal\corref{cor1}} 
\ead{abdul.awal@usask.ca}
\author[1]{Mrigank Rochan} 
\ead{mrochan@cs.usask.ca}
\author[1]{Chanchal K. Roy} 
\ead{chanchal.roy@usask.ca}
\cortext[cor1]{Corresponding author}

\affiliation[1]{organization={Department of Computer Science, University of Saskatchewan},
            addressline={110 Science Pl}, 
            city={Saskatoon},
            postcode={S7N 5C9}, 
            state={Saskatchewan},
            country={Canada}}

\begin{abstract}
Transformer-based language models of code have achieved state-of-the-art performance across a wide range of software analytics tasks, but their practical deployment remains limited due to high computational costs, slow inference speeds, and significant environmental impact. To address these challenges, recent research has increasingly explored knowledge distillation as a method for compressing a large language model of code (the teacher) into a smaller model (the student) while maintaining performance. However, the degree to which a student model deeply mimics the predictive behavior and internal representations of its teacher remains largely unexplored, as current accuracy-based evaluation provides only a surface-level view of model quality and often fails to capture more profound discrepancies in \textit{behavioral fidelity} between the teacher and student models. To address this gap, we empirically show that the student model often fails to deeply mimic the teacher model, resulting in up to \textbf{285\%} greater performance drop under adversarial attacks, which is not captured by traditional accuracy-based evaluation. In addition, to capture discrepancies in behavioral fidelity, we propose \textit{MetaCompress}, a metamorphic testing framework that systematically evaluates behavioral fidelity by comparing the outputs of teacher and student models under a set of behavior-preserving metamorphic relations. We evaluate \textit{MetaCompress} on two widely studied tasks, clone detection and vulnerability prediction, using compressed versions of popular language models of code, \textit{CodeBERT} and \textit{GraphCodeBERT}, obtained via three different knowledge distillation techniques: \textit{Compressor}, \textit{AVATAR}, and \textit{MORPH}. The results show that \textit{MetaCompress} identifies up to \textbf{62\%} behavioral discrepancies in student models, underscoring the need for behavioral fidelity evaluation within the knowledge distillation pipeline. Furthermore, the ablation study indicates that \textit{MetaCompress} is robust and effectively detects behavioral fidelity divergence, even when both the teacher and student models are assessed using transformed inputs. These results position \textit{MetaCompress} as a practical framework for evaluating the behavioral fidelity of compressed language models of code derived through knowledge distillation.
\end{abstract}

\begin{keyword}
Large Language Models of Code, Knowledge Distillation, Adversarial Attack, Metamorphic Testing, Behavioral Fidelity

\end{keyword}

\end{frontmatter}

\section{Introduction}
\label{intro}
Transformer-based \cite{vaswani2017attention} language models of code have achieved state-of-the-art performance across a wide range of software analytics tasks, including clone detection \cite{arshad2022codebert}, code summarization \cite{ahmed2024automatic}, vulnerability prediction \cite{ding2024vulnerability}, and code search \cite{chen2024code}. However, despite their impressive capabilities, the adoption of these models remains limited, particularly in consumer-grade devices (e.g., laptops) and latency-sensitive applications, due to their substantial computational demands, slow inference speed, and considerable energy consumption and carbon footprint \cite{shi2022compressing, schwartz2020green}. For instance, CodeBERT, a widely used code language model with 125 million parameters and a size of 476 MB, can incur up to 1.5 seconds of latency, consume 0.32 kWh of energy, and emit 0.14 kg of CO\textsubscript{2} on consumer-grade laptops.\footnote{When integrated into IDEs, CodeBERT may be invoked thousands of times daily by a developer, reflecting typical usage patterns \cite{hellendoorn2019code}.} To address these practical challenges, the software engineering research community has recently begun exploring knowledge distillation methods \cite{hinton2015distilling} to develop extremely compact models that reduce model size, lower inference latency, and mitigate environmental impact while maintaining task performance \cite{shi2022compressing, shi2024greening, shi2024efficient, wei2023towards, saad2024alpine, d2024compression, chen2025smaller}.

Knowledge distillation (KD) transfers knowledge from a large teacher model to a smaller student\footnote{In this study, the term student model refers to the compressed model obtained through knowledge distillation, while the term teacher model denotes the original uncompressed model.} model, maintaining comparable accuracy while requiring significantly fewer computational resources. In this context, Shi et al. \cite{shi2024greening} introduced \textit{Compressor}, a genetic algorithm-based approach that compresses code-specialized language models (e.g., CodeBERT \cite{feng2020codebert}) from 476 MB and 125 million parameters to only 3 MB with minimal loss in accuracy. Although KD effectively reduces computational costs and environmental impact, student model performance is typically evaluated using traditional accuracy-based metrics that assess whether the student model's task performance on ground-truth labels remains comparable to that of the teacher model. However, this form of evaluation offers only a surface-level view and may overlook critical discrepancies in \textbf{behavioral fidelity}—\textit{the extent to which a student model deeply mimics the predictive behavior and internal representations of its teacher counterpart}.


To investigate this limitation, we evaluate the student model’s ability to mimic the teacher by performing simple yet effective adversarial attacks \cite{yang2022natural, zhang2020generating, zeng2022extensive, tian2023code}. These attacks involve renaming identifiers and applying semantic-preserving code transformations informed by the contextual semantics of each code snippet. The empirical results reveal that although the student models achieve performance comparable to the teacher before adversarial perturbations, their performance degrades substantially under adversarial attacks, with up to \textbf{285\%} greater performance loss than the teacher model. This finding indicates that the student model does not deeply mimic the teacher’s predictive behavior and internal representations, highlighting the limitations of the traditional accuracy-based evaluation, which often overlooks more profound behavioral discrepancies between teacher and student models.

Despite these observations, current evaluation practices for knowledge distillation predominantly rely on accuracy-based metrics, which provide only a superficial assessment of model performance \cite{shi2022compressing, shi2024greening, panichella2025metamorphic}. As demonstrated in the empirical analysis, these metrics do not capture whether the student model maintains the teacher’s predictive behavior and internal representations, especially under adversarial attacks. As a result, to the best of our knowledge, there is no systematic framework for evaluating the behavioral fidelity of student models beyond task-level correctness. This limitation underscores the need for a nuanced evaluation framework that goes beyond accuracy to assess how deeply a student model mimics the predictive behavior and internal representations of its teacher, providing a more comprehensive understanding of model alignment under knowledge distillation.

To fill this gap, we introduce \textit{MetaCompress}, a testing framework inspired by metamorphic testing principles in software engineering \cite{chen2018metamorphic}, to evaluate the behavioral fidelity of the student model in knowledge distillation. In contrast to traditional accuracy-based evaluation, which compares model performance against ground-truth labels, \textit{MetaCompress} performs a head-to-head comparison of student and teacher model outputs across a set of behavior-preserving metamorphic relations, enabling a more nuanced assessment of behavioral fidelity. We design four metamorphic relations, grounded in strong empirical evidence from manual investigations, that capture discrepancies in the behavioral fidelity of student models, ranging from prediction-level disagreement to divergences in the probability distribution.

We empirically evaluate \textit{MetaCompress} on two widely studied software engineering tasks, \textit{clone detection} and \textit{vulnerability prediction}, using two popular language models of code, \textit{CodeBERT} \cite{feng2020codebert} and \textit{GraphCodeBERT} \cite{guo2020graphcodebert}. To distill their smaller student models, we apply three state-of-the-art knowledge distillation techniques designed for language models of code: \textit{Compressor} \cite{shi2022compressing}, \textit{AVATAR} \cite{shi2024greening}, and \textit{MORPH} \cite{panichella2025metamorphic}. Our extensive experiments show that \textit{MetaCompress} can reveal behavioral fidelity discrepancies of up to \textbf{62\%} across four metamorphic relations that traditional accuracy-based metrics fail to detect. Furthermore, the ablation study shows that \textit{MetaCompress} remains robust and effectively identifies behavioral fidelity divergence, even when both teacher and student models are evaluated on transformed inputs.

In summary, the main contributions of this paper are as follows: 

\begin{itemize}[leftmargin=*]
    \item \textbf{Insight:} We are the first to demonstrate that traditional accuracy-based metrics fall short in capturing behavioral fidelity between student and teacher models by measuring adversarial robustness. To address this, we reframe this evaluation challenge as a classical software testing problem and address it through the lens of metamorphic testing. Ensuring behavioral fidelity is critical because it guarantees that the student model not only provides correct outputs but also behaves consistently across scenarios, just like the teacher model, which is essential for complex tasks like coding.

    \item \textbf{Technique:} We propose \textit{MetaCompress}, a novel output-based metamorphic testing framework grounded in four behavior-preserving metamorphic relations to systematically evaluate the behavioral fidelity of student models for code

    \item \textbf{Evaluation:} We conduct a comprehensive evaluation of \textit{MetaCompress}, demonstrating its effectiveness in uncovering discrepancies in the behavioral fidelity of student models for code that are often overlooked by traditional accuracy-based metrics.
    
    \item \textbf{Open science:} Our replication package has been open-sourced for transparency and reproducibility\footnote{\url{https://doi.org/10.5281/zenodo.16127320}}.
\end{itemize}

\section{Preliminaries}

\subsection{Language Models of Code}
\label{LMCsBack}
Transformer-based \cite{vaswani2017attention} language models of code have demonstrated remarkable performance across a wide range of software analytics tasks, such as vulnerability prediction, clone detection, code summarization, code completion, code comment generation, and code search \cite{feng2020codebert, guo2020graphcodebert, lu2021codexglue, ahmad2021unified, wang2021codet5, devlin2018bert, liu2019roberta, sanh2019distilbert, husain2019codesearchnet, hou2024large}. These models learn general language representations from large amounts of unlabeled data through a two-step process: (i) pre-training and (ii) fine-tuning. During pre-training, the model employs a self-supervised objective to acquire a general understanding of source code structures and patterns. Once the training is complete, the model can be fine-tuned by updating a few parameters for specific tasks with smaller datasets, making it more efficient in terms of time and computational resources compared to training a model from scratch \cite{devlin2018bert}.

Language models for code can be broadly classified into three categories: (1) encoder-only, (2) decoder-only, and (3) encoder-decoder models \cite{zeng2022extensive}. Encoder-only models (e.g., CodeBERT \cite{feng2020codebert} and GraphCodeBERT \cite{guo2020graphcodebert}) process the entire input simultaneously and are primarily used for discriminative tasks, such as classification (e.g., vulnerability prediction and clone detection). Decoder-only models (e.g., CodeGPT \cite{lu2021codexglue}, StarCoder \cite{lozhkov2024starcoder}, and Codex \cite{chen2021evaluating}) generate output token by token and are typically employed in generative tasks, such as code generation, code completion, and summarization. Encoder-decoder models (e.g., PLBART \cite{ahmad2021unified} and CodeT5 \cite{wang2021codet5}) encode the input into a latent representation and subsequently decode it into an output sequence, making them suitable for translation-style tasks such as code summarization and programming language translation. Among these categories, encoder-only models, particularly CodeBERT and GraphCodeBERT, have been widely and effectively applied to various downstream software analytics tasks and model compression studies \cite{shi2022compressing, shi2024greening, shi2024efficient, panichella2025metamorphic, liu2026pioneer}, as evidenced by prior empirical evaluations \cite{lu2021codexglue}. Consequently, these models serve as the primary focus of our study.


\subsection{Model Compression}
\label{MCBack}
A significant barrier to the real-world deployment of language models of code is their substantial computational overhead, high memory demands, and environmental impact (e.g., carbon emissions) \cite{schwartz2020green, xu2023survey}. To mitigate these challenges and advance sustainability, a widely adopted strategy known as \textit{model compression} offers a solution that reduces the resource consumption of such models without significantly compromising performance. Model compression refers to the process of reducing a large, computationally intensive model into a smaller, more efficient version suitable for deployment on resource-constrained devices (e.g., consumer-grade laptops). In addition to deployment benefits, model compression improves inference speed and enhances overall resource efficiency. Algorithmically, the fundamental model compression techniques include: \textit{pruning} \cite{sanh2020movement}, \textit{quantization} \cite{zafrir2019q8bert}, \textit{knowledge distillation} \cite{hinton2015distilling}, \textit{neural architecture search} \cite{elsken2019neural}, \textit{weight sharing \& hashing} \cite{chen2015compressing}, and \textit{low-rank factorization} \cite{sainath2013low}-each method uniquely contributes to model optimization. Among these, pruning, quantization, and knowledge distillation (KD) have emerged as the most prominent and widely adopted approaches for compressing language models of code in recent software engineering research. \cite{shi2022compressing, shi2024greening, wei2023towards, d2024compression, chen2025smaller, liu2026pioneer, afrin2025quantization}. 

While pruning and quantization primarily remove redundant parameters or reduce numerical precision to achieve compactness, their ability to produce highly compact models is inherently limited, as aggressive compression often results in substantial degradation of representational capacity. In contrast, KD offers a distinct advantage by enabling the creation of a smaller student model that can closely approximate, and in some cases replicate, the predictive behavior of its larger teacher counterpart. Through this process, the student learns to mimic the teacher's functional behavior using soft targets such as probability distributions, internal embeddings, or logits generated by the teacher. This allows the student model to preserve the teacher's predictive behavior and generalization while maintaining a significantly reduced computational footprint. In the context of language models of code, KD has emerged as a promising compression technique for producing a deployable model that retains functional equivalence with its teacher, although the extent to which a student model truly mimics its teacher's behavior remains largely unexplored in software engineering.


\subsection{Metamorphic Testing}
\label{MTBack}
Metamorphic testing (MT) is a property-based testing paradigm widely adopted for validating software systems in the absence of a test oracle or when one is difficult to define, a challenge commonly referred to as the oracle problem \cite{chen2018metamorphic, zhang2014search}. Rather than verifying the correctness of individual outputs in isolation, MT addresses this issue by evaluating whether the program under test exhibits expected relationships between outputs across a set of well-defined metamorphic relations (MRs). An MR specifies a necessary property that captures the expected relationship between multiple inputs and their corresponding outputs. These relations are derived from the problem domain's inherent semantics and define how the output should change (or remain consistent) when the input is systematically modified. This paradigm is beneficial for testing complex systems such as machine learning models, scientific computing software, or simulation tools, where correctness is difficult to assert for individual outputs but consistency across related cases can be meaningfully assessed \cite{yang2025hallucination, xie2011testing, chen2020metamorphic, xiao2022metamorphic, segura2018metamorphic}.   

In this study, we adapt MT to evaluate how deeply a student model produced via KD mimics its teacher model's behavior. Unlike conventional MT, we do not apply transformations to the input samples. Instead, we define MRs between the teacher and student models' outputs, ensuring that when both receive the same input, their outputs exhibit a consistent relationship. These teacher–student MRs capture multiple aspects of behavioral fidelity, including agreement in predicted labels, similarity in probability distributions, and calibration consistency. By systematically evaluating these MRs, we gain a deeper understanding of behavioral fidelity between the teacher and student models beyond accuracy-based evaluation.


\subsection{Adversarial Attack}
\label{Attack}
An adversarial attack refers to the deliberate process of crafting adversarial examples by introducing subtle perturbations to the original inputs. Although these perturbed instances appear almost identical to their original counterparts from a human perspective, they can cause the model to yield erroneous or inconsistent outputs. The concept of adversarial attack was first introduced by Zegedy et al.~\cite{szegedy2013intriguing}, who demonstrated that state-of-the-art image classifiers can be misled by carefully designed pixel-level modifications that remain imperceptible to human observers. Formally, given a classifier \(f: X \rightarrow Y\) that correctly predicts the label \(y_{truth} \in Y\) for an input \(x \in X\), an adversarial attack generates a perturbed version \(x'\) such that \(f(x') \neq f(x) = y_{truth}\).

In continuous domains such as images, producing visually indistinguishable adversarial examples is relatively straightforward. However, generating adversarial examples for discrete domains like source code is considerably more complex due to the rigid syntactic and grammatical structure of programming languages~\cite{zhang2022towards, zhang2020generating}. As a result, adversarial examples designed to attack language models of code must meet three fundamental requirements. \textbf{First}, the perturbation must effectively mislead the model and reduce its predictive performance. \textbf{Second}, the perturbed code must remain syntactically correct, adhering to the formal rules of the programming language; for example, in Python, identifiers cannot begin with digits and may only contain letters, digits, or underscores. \textbf{Third}, the perturbed code \(x'\) must preserve the semantics of the original code \(x\), maintaining identical functionality and behavior when executed under the same conditions. When an input instance is modified in accordance with these constraints to generate a perturbed version, the resulting instance constitutes an adversarial example; if it causes the model to produce an incorrect or inconsistent output, it is regarded as an adversarial attack.

\subsection{Adversarial Robustness}
\label{Robust}
Adversarial robustness describes a model's capacity to maintain its predictive performance and stability when exposed to adversarial perturbations. A model is considered adversarially robust if it continues to generate consistent, accurate outputs despite deliberate input modifications intended to mislead it. In the context of machine learning, this concept reflects a model's resilience to adversarial attacks that exploit its sensitivity to small perturbations in the input space \cite{szegedy2013intriguing, zhang2022towards}.

\subsection{Kullback–Leibler Divergence}
\label{KLDiv}
The Kullback-Leibler (KL) divergence \cite{kullback1951information} is a measure of how one probability distribution \( Q \) diverges from a reference probability distribution \( P \). For discrete distributions defined over the same support, it is mathematically expressed as:

\begin{equation}
    D_{\mathrm{KL}}(P \parallel Q) = \sum_{i} P(i) \log \frac{P(i)}{Q(i)}
    \label{KLDivEqn}
\end{equation}


where \( P(i) \) and \( Q(i) \) represent the probabilities of the \( i \)-th event under distributions \( P \) and \( Q \), respectively. KL divergence is always non-negative and equals zero if and only if \( P = Q \). It is commonly used in machine learning to quantify the difference between predicted and target distributions, particularly in tasks such as knowledge distillation. An important characteristic of KL divergence is that it is \textit{not symmetric}, meaning that in general, $ D_{\mathrm{KL}}(P \parallel Q) \neq D_{\mathrm{KL}}(Q \parallel P)$.




\section{Empirical Study}
\label{MimicTest}
This section presents the design of our empirical study, which aims to evaluate whether the student model deeply mimics the teacher’s predictive behavior and internal representations during knowledge distillation by assessing its adversarial robustness. We hypothesize that if the student model deeply mimics the teacher during knowledge distillation, it should exhibit adversarial performance comparable to the teacher's. Any substantial deviation in robustness would indicate that the student model fails to deeply mimic the teacher’s predictive behavior and internal representations. We first describe the selected knowledge distillation approaches for code-language models, followed by details on datasets, models, adversarial attack approaches, and the experimental setup. Finally, we outline the evaluation metrics used in the analysis and present the corresponding results in the following section.

\subsection{Study Design}
\label{MimicStudyDesign}

\subsubsection{Knowledge Distillation Approaches}
\label{KDApproaches}
In recent years, several knowledge distillation (KD) approaches have been proposed for language models of code, including \textit{Compressor} \cite{shi2022compressing}, \textit{AVATAR} \cite{shi2024greening}, \textit{SODA} \cite{chen2025smaller}, \textit{MORPH} \cite{panichella2025metamorphic}, and \textit{PIONEER} \cite{liu2026pioneer}. In this study, we select \textit{Compressor}, \textit{AVATAR}, and \textit{MORPH} as our KD approaches to generate student models, as the replication package for \textit{SODA} is not fully available\footnote{https://shorturl.at/Se0f7}. Furthermore, both \textit{MORPH} and \textit{PIONEER} incorporate robustness as an additional objective of knowledge distillation; however, because \textit{MORPH} optimizes a broader set of objectives than \textit{PIONEER}, we exclude \textit{PIONEER} from our empirical investigation.

\textit{Compressor}, \textit{AVATAR}, and \textit{MORPH} generate deployable models from language models of code using neural architecture search guided by multi-objective optimization. \textit{Compressor} explores various architecture-related hyperparameters, including the \textit{number of network layers}, \textit{dimensionality of network layers}, \textit{number of attention heads}, \textit{dimensionality of the feed-forward layers}, and \textit{vocabulary size}. Whereas, both \textit{AVATAR} and \textit{MORPH} use the same configuration space for distillation. The configuration parameters include the choice of \textit{tokenizer} (\textit{BPE}, \textit{WordPiece}, \textit{Unigram}, or \textit{Word}), a \textit{vocabulary size} ranging from 1,000 to 46,000, and \textit{number of hidden layers} ranging from 1 to 12. \textit{Hidden layer dimension} varies from 16 to 256, with \textit{activation functions} selected from \textit{gelu}, \textit{relu}, \textit{silu}, and \textit{gelu\_new}. \textit{Hidden and attention dropout probabilities} are each set to a value between 0.2 and 0.5. \textit{Intermediate layer size} ranges from 32 to 3,072, and \textit{number of attention heads} ranges from 1 to 12. \textit{Maximum sequence length} is configured between 256 and 512, while \textit{position embedding type} can be \textit{absolute}, \textit{relative\_key}, or \textit{relative\_key\_query}.

\textit{Compressor} uses the genetic algorithm to identify a simplified architecture that maximizes \textit{giga floating-point operations} (GFLOPs) while minimizing \textit{model size} within a fixed memory budget. \textit{AVATAR} compresses language models of code into deployable versions by jointly optimizing \textit{model size}, \textit{inference latency}, \textit{energy consumption}, and \textit{accuracy}. It formulates the problem as a multi-objective configuration tuning task, using an SMT solver to define the search space and a tailored optimization algorithm to identify a Pareto-optimal set of configurations for the student model. Based on the \textit{AVATAR}, \textit{MORPH} considers \textit{metamorphic robustness} a crucial property for multi-objective optimization in architecture selection for the student model. In addition, it distills knowledge from the teacher model not only on the soft targets but also on the metamorphic code's outputs, assuming the teacher model always produces correct outputs for the metamorphic code. Subsequently, knowledge distillation is performed by minimizing the loss between the soft targets produced by the teacher model and those generated by the student model. For example, \textit{Compressor} employs the KL divergence loss function, defined as follows:

\begin{equation}
    \mathcal{L} = -\frac{1}{n} \sum_{i}^{n} softmax(\frac{p_i}{T})\text{log} \Bigl( softmax(\frac{q_i}{T}) \Bigl) T^2
    \label{KLDLoss}
\end{equation}

In ``\eqref{KLDLoss}'', $p_i$ and $q_i$ represent the teacher and student model logits, respectively, with $T$ denoting the temperature parameter controlling the softmax distribution \cite{hinton2015distilling}. While the teacher logits remain fixed, the student logits are updated throughout the training process. It is essential to note that we utilize \textit{off-the-shelf} compressed versions of CodeBERT and GraphCodeBERT, produced by \textit{MORPH} and \textit{AVATAR}, for clone detection and vulnerability prediction tasks in this empirical evaluation.

\subsubsection{Datasets and Subjects}
\label{DataModels}
To minimize experimental bias in evaluating whether the student model deeply mimics the teacher’s predictive behavior and internal representations during knowledge distillation, we adopt the same models and datasets used in the knowledge distillation–based approaches described in Section \ref{KDApproaches}. Specifically, we select CodeBERT \cite{feng2020codebert} and GraphCodeBERT \cite{guo2020graphcodebert} models and two well-studied downstream tasks: clone detection and vulnerability prediction, which serve as standard benchmarks in knowledge distillation research for software engineering \cite{shi2022compressing, shi2024greening, shi2024efficient, panichella2025metamorphic, liu2026pioneer}.

\textbf{Clone Detection (CD):} This task aims to determine whether two given code snippets are syntactically or semantically equivalent. To streamline the fine-tuning process, we adopt the filtered BigCloneBench dataset \cite{svajlenko2014towards} provided by Wang et al. \cite{wang2020detecting}. The final dataset includes 90,102 training examples, along with 4,000 examples each for validation and testing. 

\textbf{Vulnerability Prediction (VP):} This task aims to determine whether a given code snippet contains a vulnerability. We conduct our experiments using the Devign\footnote{\url{https://sites.google.com/view/devign}} dataset, introduced by Zhou et al. \cite{zhou2019devign}. The dataset comprises 27,318 functions sourced from two widely used open-source C libraries, FFmpeg and Qemu. Each function is manually annotated to indicate whether it contains vulnerabilities. We use the CodeXGLUE benchmark splits \cite{lu2021codexglue}, including training, development, and test sets with 27,318, 2,732, and 2,732 examples, respectively.

\textbf{CodeBERT} is a pre-trained language model that shares the same architecture and a similar pre-training strategy as RoBERTa \cite{liu2019roberta}. It consists of twelve Transformer layers, each with 12-headed multi-head attention mechanisms and a hidden size of 768. CodeBERT is pre-trained using two objectives: \textit{masked language modeling}, which predicts masked tokens, and \textit {replaced token detection}, which identifies substituted tokens in the input. The model is trained on bimodal data comprising natural language–code pairs across six programming languages from the CodeSearchNet dataset \cite{husain2019codesearchnet}. 

\textbf{GraphCodeBERT} shares CodeBERT's architecture but incorporates data-flow information from source code during pre-training. Additionally, it introduces a pre-training task that predicts whether a data flow edge exists between two nodes.

\subsubsection{Adversarial Attack Approaches}
\label{AdvAttackApproach}
In this study, adversarial attacks are employed to evaluate how deeply the student model mimics the teacher’s predictive behavior and internal representations during knowledge distillation. Adversarial attacks are typically categorized as black-box or white-box, depending on the attacker’s level of access to the model. Following the recommendation of Du et al. \cite{du2023extensive}, we adopt black-box attacks, as they rely solely on model outputs and are widely used for evaluating language models of code \cite{yang2022natural, zhang2020generating}. The perturbations introduced through these attacks are deliberately subtle to maintain the syntactic and semantic integrity of the source code, as discussed in Section \ref{Attack}, while inducing mispredictions. To this end, we employ three established black-box techniques: ALERT \cite{yang2022natural}, Metropolis Hastings Modifier (MHM) \cite{zhang2020generating}, and WIR-Random \cite{zeng2022extensive}, each of which performs identifier substitution to assess model robustness without altering code functionality. To further enhance robustness evaluation, we also employ \textbf{CODA} \cite{tian2023code}, a black-box attack technique that combines identifier renaming with semantic-preserving code transformations. This enables assessment of model robustness under a wider range of functionally equivalent program variations.

\textbf{ALERT} considers both operational and natural semantics, using greedy and genetic algorithms to select substitutions, with 30 candidates per identifier. \textbf{MHM} utilizes Metropolis-Hastings sampling for iterative identifier substitution, with a maximum of 100 iterations and 30 candidates per iteration. \textbf{WIR-Random} ranks identifiers based on their impact on model predictions and replaces them sequentially with randomly chosen candidates. In contrast to these identifier-centric techniques, \textbf{CODA} utilizes code differences between a target input and reference inputs with divergent predictions to guide adversarial example generation. It combines identifier renaming with semantic-preserving structural transformations. By considering both the identifier level and structural differences, CODA generates functionally equivalent yet more diverse adversarial examples.

It is important to note that only instances correctly classified before perturbation are included, ensuring that the robustness evaluation is based exclusively on valid predictions and that any observed degradation in model performance is caused solely by adversarial perturbations rather than pre-existing misclassifications \cite{yang2022natural, zhang2020generating}.

\subsubsection{Experimental Settings}
\label{ExpSettings}
We fine-tune CodeBERT using the hyperparameter settings and training pipeline provided by the CodeXGLUE benchmark \cite{lu2021codexglue}, while for GraphCodeBERT, we adopt the settings specified in its original paper \cite{guo2020graphcodebert}. To ensure consistency across our experiments, we adopt the data-splitting strategy used in knowledge distillation–based approaches such as \textit{MORPH} and \textit{AVATAR}, which divide the training data into two equal halves: one for fine-tuning and the other for knowledge distillation. We apply the same fine-tuning split to all models. Despite being trained on only half of the original dataset, all models achieve performance comparable to the state-of-the-art results reported in the CodeXGLUE benchmark \cite{lu2021codexglue}, the GraphCodeBERT paper \cite{guo2020graphcodebert}, and knowledge distillation–based compression studies such as \textit{Compressor}, \textit{AVATAR}, and \textit{MORPH}.

\subsubsection{Evaluation Metrics}
\label{EvalAdvMetric}
Following the evaluation protocols of CodeXGLUE \cite{lu2021codexglue}, Yang et al. \cite{yang2022natural}, and Shi et al. \cite{shi2022compressing, shi2024greening}, we use \textbf{accuracy} as the primary metric to evaluate model performance on downstream tasks.

To evaluate the performance of both teacher and student models under adversarial conditions, we measure the effectiveness of the applied attack methods. A higher attack effectiveness indicates greater model vulnerability to adversarial perturbations. We employ the \textit{Attack Success Rate} (ASR) metric \cite{yang2022natural} to measure the effectiveness of an attack method. ASR measures the proportion of adversarial samples for which the model fails to produce a correct prediction after the attack and is formally defined as follows:

\begin{equation}
    \text{ASR} = \frac{|\{x \in X \wedge M(x') \neq M(x)\}|}{\left|\left\{x \in X \mid M(x)=y\right\}\right|}
    \label{ASREQN}
\end{equation}

Where $x$ is an original input, $x'$ is a modified input. A higher ASR value indicates greater attack effectiveness, reflecting the models' adversarial vulnerabilities.

The robustness of compressed models to adversarial attacks depends strongly on the quality of the adversarial examples generated. If these examples contain syntactic or semantic errors, model mispredictions may not accurately reflect a lack of robustness but rather the model’s inability to handle unrealistic inputs. Moreover, excessive modifications to the original code during adversarial example generation can lead to out-of-distribution instances, thereby reducing the reliability of robustness assessments. To ensure a fair and meaningful evaluation, we conduct a qualitative analysis to assess the effectiveness of the generated adversarial examples and to verify that the applied perturbations are minimal yet sufficient to mislead the models. Because the inputs for the selected software analytics tasks are source code, we employ four established metrics from Du et al. \cite{du2023extensive}: \textit{Identifier Change Rate} (ICR), \textit{Token Change Rate} (TCR), \textit{Average Edit Distance} (AED), and \textit{Average Code Similarity} (ACS), to measure the quality of the adversarial examples.

For a set of $m$ adversarial examples, each $i$-th code snippet contains $k_i$ identifiers, among which $n_i$ are modified to generate the adversarial example. The \textbf{ICR} is computed as follows:

\begin{equation}
\text{ICR} = \frac{\sum_{1}^{m} n_i}{\sum_{1}^{m} k_i}
\label{ICREQN}
\end{equation}

The \textbf{TCR} represents the proportion of modified tokens relative to the total number of tokens in the original code. The \textbf{AED} quantifies the average number of character-level edits required to transform one token into another, reflecting the textual intensity of the perturbation. The \textbf{ACS} measures cosine similarity between embeddings produced by CodeBERT before and after the adversarial modifications. In general, high-quality adversarial examples are characterized by lower ICR, TCR, and AED values, along with higher ACS scores, indicating that the perturbations are subtle yet effective while preserving code correctness.

\subsection{Empirical Results}
\label{MimicStudyResults}
This section presents the empirical results of our study designed to address the first research question, \textbf{RQ1}: \textit{Does the student model deeply mimic the teacher’s predictive behavior and internal representations during knowledge distillation?}

We address this research question by evaluating the performance of student models under adversarial attacks relative to their teacher counterparts. In addition, we examine whether traditional accuracy-based evaluation can capture discrepancies in behavioral fidelity arising from the student models’ inability to deeply mimic their teacher models. Since \textit{AVATAR} outperforms \textit{Compressor} and \textit{MORPH} considers metamorphic robustness during knowledge distillation, we compare the adversarial robustness of student models compressed by \textit{AVATAR} and \textit{MORPH} with that of the teacher model for this empirical analysis.

To answer this research question, we first compare the performance of the teacher and student models on the original, unperturbed test samples using \textit{accuracy} as the evaluation metric. Table \ref{AccTab} presents the accuracy comparison between teacher and student models across the clone detection and vulnerability prediction tasks. Overall, the results indicate that the student models maintain performance levels comparable to their teacher counterparts, with accuracy differences remaining within a narrow margin (mostly below 3\%). For instance, in clone detection, the student models CodeBERT and GraphCodeBERT compressed by \textit{AVATAR} and \textit{MORPH} achieve accuracies close to those of their teacher models, with only minor reductions observed for \textit{AVATAR}. A similar trend is observed in vulnerability prediction, where student models exhibit minimal degradation and, in some cases, nearly identical performance to the teachers.


\begin{table}[ht]
\centering
\caption{Accuracy (\%) comparison between teacher (uncompressed) and student (compressed) models across clone detection and vulnerability prediction tasks.}
\label{tab:accuracy_comparison}
\resizebox{\columnwidth}{!}{
\begin{tabular}{lccccc}
\hline
\textbf{Task} & \textbf{Model} & \textbf{Teacher} & \textbf{\textit{Compressor}} & \textbf{\textit{AVATAR}} & \textbf{\textit{MORPH}} \\
\hline
Clone Detection & CodeBERT & 96.7 & 95.5 & 93.1 & 95.7 \\
                & GraphCodeBERT & 96.5 & 93.9 & 93.6 & 96.3 \\ 
\hline
Vulnerability Prediction & CodeBERT & 61.7 & 59.2 & 60.6 & 59.4 \\
                         & GraphCodeBERT & 60.8 & 60.3 & 59.3 & 59.7 \\
\hline
\end{tabular}
}
\label{AccTab}
\end{table}

Although the student models achieve task performance comparable to that of their teachers, these results provide only a surface-level view of knowledge transfer. Accuracy-based evaluation captures output correctness but may overlook discrepancies in behavioral fidelity between teacher and student models. Two models may achieve similar accuracy yet differ in representation, generalization, or response to perturbations. Hence, accuracy alone may not reveal whether the student model deeply mimics the teacher’s predictive behavior and internal representations. To address this, we further examine both models under adversarial attack scenarios to gain deeper insight into their behavioral fidelity.



\begin{figure}[htbp]
\centering
\includegraphics[width=\textwidth,height=0.7\textheight,keepaspectratio]{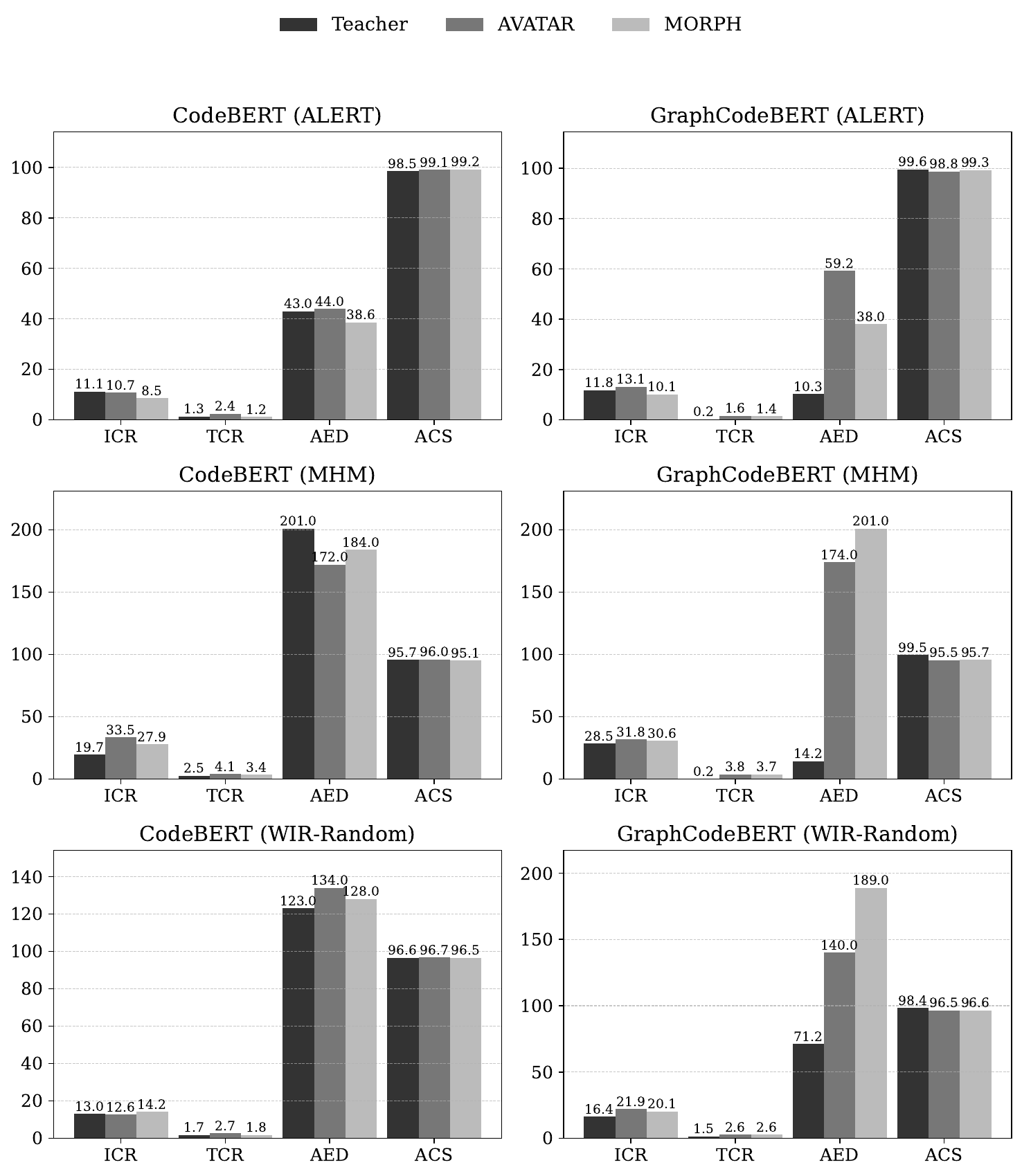}
\caption{Quality analysis of adversarial examples across various models and evaluation metrics.}
\label{AdvExampleQuality}
\end{figure}

Before analyzing model robustness under adversarial conditions, we first evaluate the quality of the adversarial examples generated by each attack technique. Excessive or unrealistic modifications can distort the syntactic or semantic structure of the original code, leading to misleading robustness assessments. To ensure reliability, we examine the syntactic and semantic integrity of the adversarial examples produced by each attack technique, confirming that the applied perturbations are minimal yet effective. This analysis uses the CodeBERT and GraphCodeBERT models trained on the clone detection task, with complete results for the vulnerability prediction task available in our replication package. Since all attack techniques modify only \textit{identifiers}, the resulting code retains its original semantics. Figure \ref{AdvExampleQuality} illustrates the quality assessment of adversarial examples generated by different attack techniques against both teacher and student models.

 Overall, the results indicate that these attack techniques effectively preserve the syntactic integrity of the generated adversarial examples. Specifically, the examples consistently exhibit high ACS values and low TCR scores across both CodeBERT and GraphCodeBERT models, demonstrating minimal syntactical distortion. The adversarial examples also maintain strong overall quality across the AED and ICR metrics for both the teacher and student models. Although a slight degradation is observed in certain cases—for instance, the MHM attack shows comparatively higher AED values—the overall differences remain limited. We further perform the Friedman test \cite{friedman1937use} to determine whether the three attack techniques (ALERT, MHM, and WIR-Random) generate adversarial examples of significantly different quality for each model variant of CodeBERT and GraphCodeBERT. A $p$-value below 0.05 indicates that the differences in the quality of adversarial examples produced by the three attack techniques are statistically significant. However, the experimental results show no statistically significant differences among the attack methods for either CodeBERT (\(p = 0.47\), \(p = 0.37\), and \(p = 0.37\) for Teacher, \textit{AVATAR}, and \textit{MORPH}, respectively) or GraphCodeBERT (\(p = 0.63\), \(p = 0.37\), and \(p = 0.37\) for Teacher, \textit{AVATAR}, and \textit{MORPH}, respectively). These results indicate that all three attack techniques generate adversarial examples of comparable quality across the evaluated metrics (ICR, TCR, AED, and ACS).

Although the preceding results show that the adversarial examples essentially preserve syntactic and semantic quality across both teacher and student models, it remains important to examine whether any subtle quality differences exist between these two model categories. To investigate this, we compare the quality of adversarial examples generated by each attack technique for the teacher and student models across all the evaluation metrics. The Wilcoxon signed-rank test \cite{woolson2007wilcoxon} is used to assess the statistical significance of these differences. The comparison between the Teacher and \textit{AVATAR} models (\(p = 0.13\)) and between the Teacher and \textit{MORPH} models (\(p = 0.85\)) indicates that the quality of adversarial examples remains statistically comparable, with no significant differences observed in either case. These findings suggest that the attack techniques generate adversarial examples of similar syntactic and semantic integrity across both teacher and student models, confirming that the subsequent robustness analysis is based on adversarial inputs of consistent quality and that any observed performance degradation can be attributed to the models’ inability to maintain robustness against adversarial perturbations.

\begin{figure}[htbp]
\centering
\includegraphics[width=\textwidth,height=0.7\textheight,keepaspectratio]{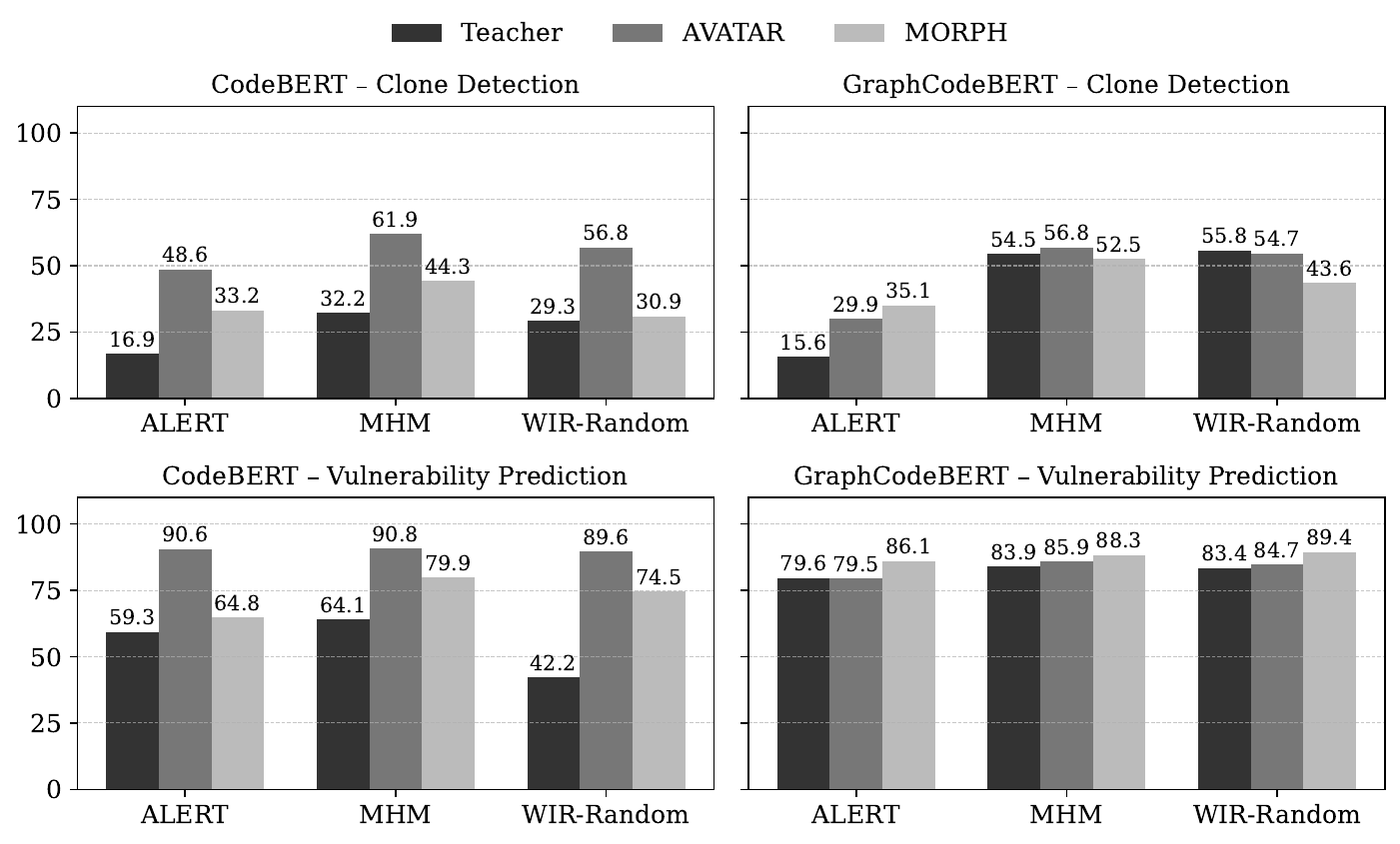}
\caption{Attack success rate (ASR) metric values across identifier-renaming adversarial attack techniques, models, tasks, and knowledge distillation methods.}
\label{ASRCompare}
\end{figure}

Having rigorously evaluated the quality of the generated adversarial examples, we next examine how effectively the teacher and student models withstand adversarial attacks. Figure \ref{ASRCompare} presents the Attack Success Rate (ASR) values for the teacher and student models (\textit{AVATAR} and \textit{MORPH}) across the ALERT, MHM, and WIR-Random attack techniques for both the clone detection and vulnerability prediction tasks. Overall, the results indicate that student models produced through knowledge distillation exhibit substantially lower robustness than their corresponding teacher models, as evidenced by consistently higher ASR values across most attack settings, with only two exceptions. In the clone detection task, the teacher model demonstrates lower robustness than the \textit{MORPH} student model under the MHM and WIR-Random attacks, and than the \textit{AVATAR} student model under the MHM attack. Although the robustness differences between the teacher and student models are relatively small for the GraphCodeBERT model and the vulnerability prediction task, they are considerably larger in all other experimental settings. For instance, in the clone detection task with the CodeBERT model under the ALERT attack, the \textit{AVATAR} student model experiences approximately three times greater performance degradation than the teacher model.

\begin{figure}[htbp]
\centering
\includegraphics[width=\textwidth,height=0.7\textheight,keepaspectratio]{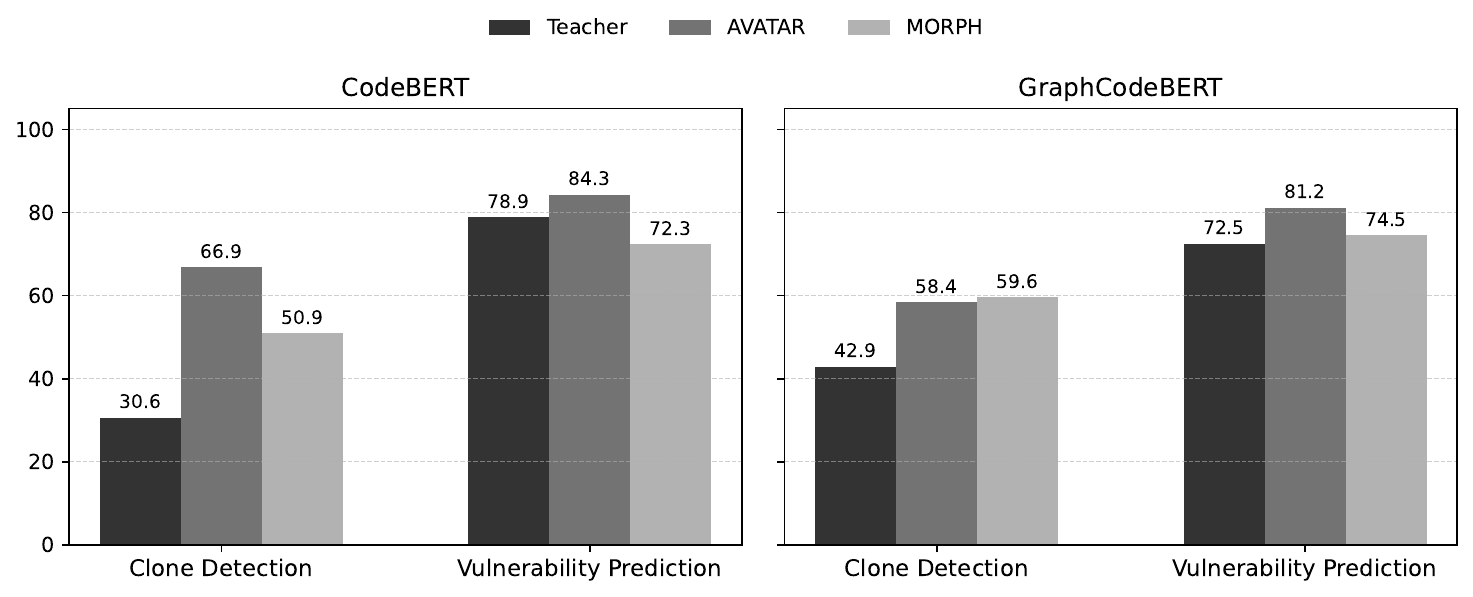}
\caption{Attack success rate (ASR) metric values across CODA adversarial attack, models, tasks, and knowledge distillation methods.}
\label{ASRCODACompare}
\end{figure}

To further assess the robustness of teacher and student models to a broader class of semantic-preserving adversarial perturbations, we conduct additional experiments using the CODA attack. Figure \ref{ASRCODACompare} presents the Attack Success Rate values under CODA. The results align with the trends observed for ALERT, MHM, and WIR-Random attacks. Across both tasks and model architectures, student models trained via knowledge distillation generally achieve higher ASR values than their teacher counterparts, indicating reduced robustness to CODA-induced perturbations. This finding applies to both AVATAR and MORPH students, confirming that the previously observed robustness degradation is not restricted to identifier-level attacks.

Taken together, these findings suggest that student models distilled from their teachers, particularly those produced by \textit{AVATAR} and, to a lesser extent, by \textit{MORPH}, are more susceptible to adversarial attacks than their teacher counterparts. Surprisingly, although \textit{MORPH} generates student models by incorporating metamorphic robustness during knowledge distillation, the resulting student models exhibit lower robustness than those produced by \textit{AVATAR} across all the attack techniques for the GraphCodeBERT model in the vulnerability prediction task. The results further reveal that student models fail to deeply mimic the teacher’s predictive behavior and internal representations. Although knowledge distillation preserves task accuracy under clean inputs, it weakens the model’s ability to generate stable and consistent predictions when exposed to adversarial perturbations. As a result, traditional accuracy-based evaluation cannot capture these discrepancies in behavioral fidelity, which become evident only under adversarial conditions. These observations highlight the need for more comprehensive evaluation frameworks that assess behavioral fidelity beyond accuracy.

\begin{center}
\begin{tcolorbox}[
    enhanced,
    attach boxed title to top left={yshift=-3mm,yshifttext=-1mm}, 
    colback=mycolor_box,                 
    colframe=black,                
    colbacktitle= mycolor_title,            
    coltitle=black,                
    title=Answer to RQ1,            
    fonttitle=\bfseries,           
    boxed title style={size=small},
]

Since the adversarial examples across all teacher and student models exhibit comparable syntactic and semantic quality, the observed robustness gap can be attributed to the student models’ inability to deeply mimic the teacher models’ predictive behavior and internal representations, a limitation that traditional accuracy-based evaluation fails to reveal.

\end{tcolorbox}
\label{RQ1Result}
\end{center}

\section{Approach: \textit{MetaCompress} Framework}
\label{FrameworkDesign}
\begin{figure}[htbp]
  \centering
  \includegraphics[width=12cm, height=5.75cm]{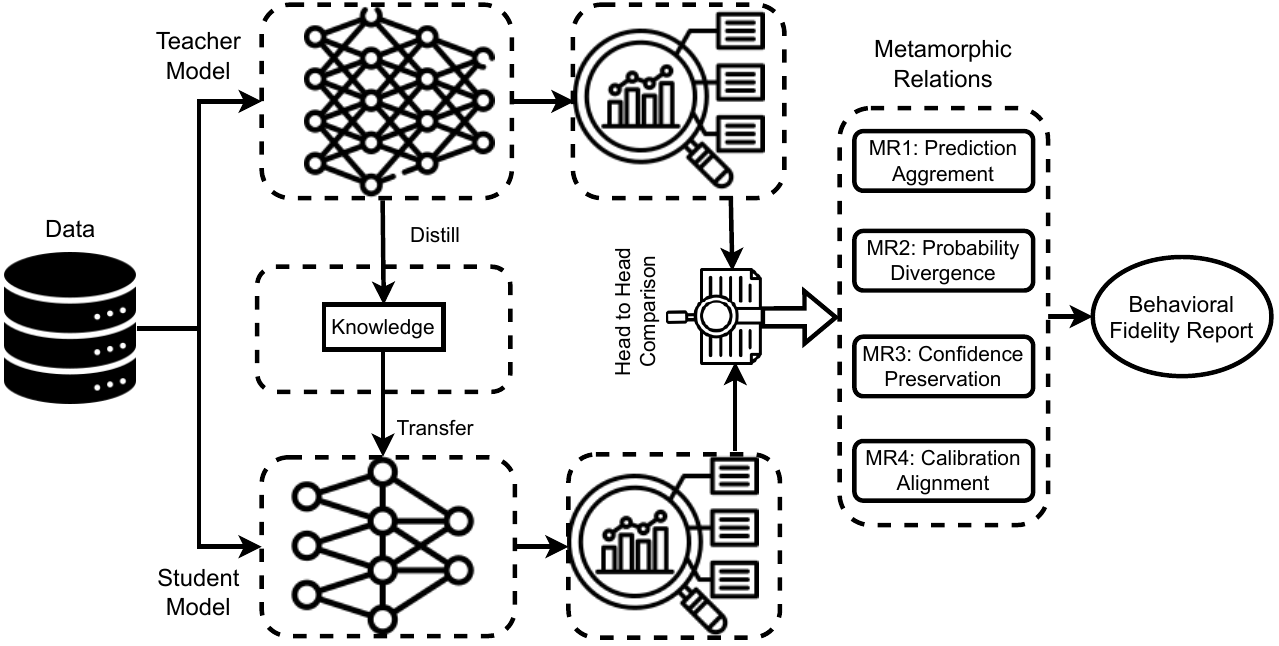}
  \caption{Workflow of the \textit{MetaCompress} framework, comparing teacher and student model outputs under behavior-preserving metamorphic relations to assess behavioral fidelity.}
  \label{MetaCompressDiagram}
\end{figure}

The empirical findings in Section~\ref{MimicStudyResults} show that traditional accuracy-based evaluation is insufficient to capture behavioral discrepancies between teacher and student models, which become evident under adversarial conditions due to the student models’ inability to deeply mimic their teacher counterparts' predictive behavior and internal representations. Motivated by this observation, we introduce \textit{MetaCompress}, a metamorphic testing framework for evaluating the behavioral fidelity of compressed models by framing the task as a classical software testing problem and applying a metamorphic testing strategy. The framework identifies key behavioral invariants that should be preserved after knowledge distillation and formalizes a set of output-based metamorphic relations that serve as oracle-free tests to assess behavioral fidelity between teacher and student models. Figure \ref{MetaCompressDiagram} presents the overall design and workflow of the proposed \textit{MetaCompress} framework.

\subsection{Behavioral Fidelity as a Software Testing Problem}
\label{ProblemFormulation}
Knowledge distillation transfers knowledge from a large teacher model to a smaller student model, aiming to preserve the student’s performance on the original task. However, our empirical results demonstrate that the student model does not deeply mimic the teacher’s internal behavior, raising an important question: \textit{Does the student model behave consistently with its teacher?} This concern closely parallels a classic problem in software engineering: verifying that a modified system preserves the intended behavior of its original version. In this context, we treat the student model as a transformed version of the teacher model and frame the evaluation of behavioral fidelity as a software testing problem. From this perspective, the teacher and student models represent two versions of the same system under test (SUT), and our goal is to assess whether behavioral fidelity is maintained after knowledge distillation. Formally, let \( M: \mathcal{X} \rightarrow \mathcal{Y} \) denote the teacher model and \( M': \mathcal{X} \rightarrow \mathcal{Y'} \) its student variant. Given a set of test inputs \( \mathcal{T} \subseteq \mathcal{X} \) and a task-specific equivalence relation \( \equiv \subseteq \mathcal{Y} \times \mathcal{Y'} \), the problem of behavioral fidelity testing is to determine whether:

\begin{equation}
\forall x \in \mathcal{T}, \quad M(x) \equiv M'(x)
\end{equation}
That is, whether the student model produces outputs that are behaviorally consistent with those of the teacher model across all test cases.

In traditional software engineering, techniques such as regression testing \cite{wong1997study}, mutation testing \cite{jia2010analysis}, and metamorphic testing \cite{chen2018metamorphic} are commonly employed to assess the reliability of modified systems. However, evaluating language models of code presents unique challenges compared to traditional software systems. For example, traditional software typically produces deterministic outputs, while language models of code yield probabilistic outputs, and knowledge distillation may alter output distributions even when top-1 predictions remain unchanged. Additionally, we aim to assess the behavioral fidelity of a student model compared to its teacher counterpart. Although ground-truth labels are available for tasks such as clone detection or vulnerability prediction, our objective is not to directly compare the outputs of the two models against the ground-truth.  Instead, we compare their outputs after model execution, treating the teacher model as the behavioral reference. Notably, the teacher model's behavior is not known until it is executed on test data. We refer to this scenario as an instance of the \textit{oracle problem}, where the absence of a definitive oracle hinders straightforward correctness verification.

As a result, conventional software testing techniques such as regression testing, which depend on exact output matches, are often too rigid—minor differences in probabilistic outputs or generated sequences do not necessarily indicate functional errors. Mutation testing is also ill-suited, as neither model is inherently faulty. In contrast, metamorphic testing provides a principled approach to overcoming the oracle problem by defining expected relations (as detailed in Section \ref{MetamorphicRelationsDesign}) between the outputs of the teacher and student models, without requiring explicit ground-truth. These relations act as relational oracles, enabling us to verify whether the teacher and student models produce consistent outputs to the same inputs. Finally, metamorphic testing has been widely adopted for evaluating machine learning–based software systems and large language models \cite{xie2020mettle, sun2024fairness, kwan2024mt, zhang2020machine, wu2025detecting, yang2025hallucination}, demonstrating its versatility and effectiveness in oracle-free evaluation scenarios. Building on this foundation, \textit{MetaCompress} adopts a metamorphic testing–based approach to evaluate behavioral fidelity of student models, framing it as an oracle-free testing problem.

\subsection{Designing Output-Based Metamorphic Relations}
\label{MetamorphicRelationsDesign}
\textit{MetaCompress} is designed to evaluate discrepancies in behavioral fidelity between teacher and student models. Without a head-to-head comparison of teacher and student models on identical inputs, it is difficult to determine whether the student model maintains the teacher model's behavioral fidelity. For example, a manual inspection of a student GraphCodeBERT model generated using \textit{Compressor} for the vulnerability prediction task illustrates this limitation. Figure 3(a) shows a case in which both models predict the same class, yet their output probability distributions diverge substantially, with a Kullback–Leibler divergence of 0.6177. In contrast, as shown in Figure 3(b), the teacher and student models yield different class predictions for the same input, revealing behavioral differences that accuracy alone fails to capture. To systematically assess such differences, \textit{MetaCompress} defines a set of output-based metamorphic relations that encode expected behavioral invariants, such as class agreement and similarity in probability distributions, between teacher and student models. Unlike traditional metamorphic testing, which typically applies transformations to inputs, these relations operate directly on model outputs, enabling a structured and quantitative evaluation of behavioral fidelity. The following subsections describe the design rationale, empirical motivation, and evaluation metrics for each metamorphic relation incorporated into the \textit{MetaCompress} framework.

\begin{figure}[htbp]
  \centering
  \includegraphics[width=7.5cm, height=8cm]{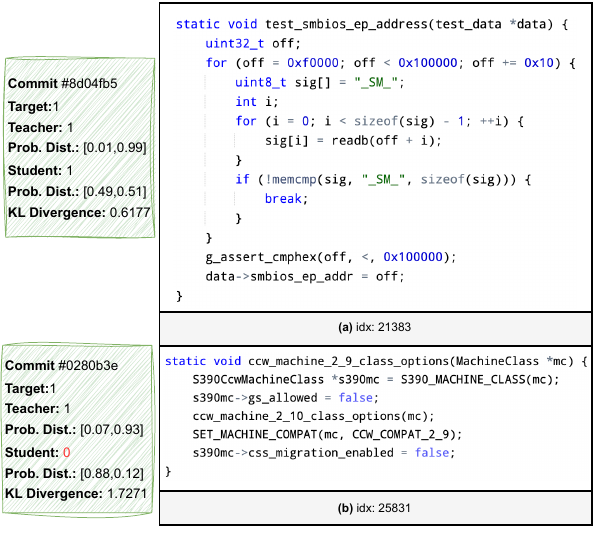}
  \caption{Behavioral fidelity discrepancies between the outputs of teacher and student models.}
  \label{MotivatingExample}
\end{figure}

\subsubsection{MR1 - Prediction Agreement}
\label{MR1Label}
\paragraph*{\textbf{Description}} Given a set of test inputs $X={x_1, x_2, ..., x_n}$, a teacher model $M_T$ and a student model $M_S$, for each input $x_i \epsilon X$, the prediction of $M_S$ should match of $M_T$. We define \textbf{MR1} as follows:

\begin{equation}
    M_S(x_i)=M_T(x_i)
    \label{MR1Equation1}
\end{equation}

\paragraph*{\textbf{Rationale}} This MR captures the most intuitive form of behavioral fidelity: \textit{top-1} label consistency. Our empirical observations (as shown in Figure 3(b)) suggest that prediction disagreement may be associated with divergence in model reasoning. Therefore, \textbf{MR1} serves as a meaningful indicator of behavioral fidelity.

\paragraph*{\textbf{Evaluation metric}} To measure this MR, we consider the \textit{Label Loyalty} metric that quantifies the degree to which this MR holds across all the test instances. It computes the percentage of instances for which the student model’s prediction matches the teacher model’s prediction, analogous to accuracy:

\begin{equation}
    \text{Loyalty}_{\text{lab}}=\frac{1}{n} \sum_{i=1}^{n}1[M_S(x_i)=M_T(x_i)]
    \label{MR1Equation2}
\end{equation}

Thus, $\text{Loyalty}_{\text{lab}}$ quantitatively measures the success rate of this MR. A higher metric value indicates greater behavioral fidelity preservation in the student model.

\subsubsection{MR2 - Probability Distribution Similarity}
\label{MR2Label}
\paragraph*{\textbf{Description}} This MR assesses the similarity of output probability distributions between the two models by evaluating whether, for a given input $x$, the predicted distribution of $M_S$ closely aligns with that of $M_T$. Given a set of test inputs $X={x_1, x_2, ..., x_n}$, let \( P_T(y \mid x) \) and \( P_S(y \mid x) \) represent the probability distributions over the output space \( \mathcal{Y} \) produced by \( M_T \) and \( M_S \), respectively. We define \textbf{MR2} as follows:

\begin{equation}
    \quad D\left(P_T(\cdot \mid x), P_S(\cdot \mid x)\right) \leq \delta, \quad \forall x \in {X}
    \label{MR2Equation1}
\end{equation}

Where \( D(\cdot, \cdot) \) is a divergence metric (e.g., Kullback–Leibler divergence), and \( \delta \geq 0 \) is a predefined tolerance threshold.

\paragraph*{\textbf{Rationale}} Although the teacher and student models may produce the same predicted label, their underlying probability distributions can differ substantially. Our empirical observations (Figure 3(a)) reveal such differences, as measured by the Kullback–Leibler divergence metric (e.g., 0.6177), suggesting that prediction agreement (\textbf{MR1}) alone is insufficient to ensure behavioral fidelity.

\paragraph*{\textbf{Evaluation metric}} To quantify the difference between the two probability distributions, we use the Kullback–Leibler divergence, as defined in ``\eqref{KLDivEqn}''. Given a tolerance $\delta=0.5$ (Inspired by sigmoid-based methods for binary classification \cite{lecun2015deep}), we then count the test instances for which MR2 holds:

\begin{equation}
\text{Loyalty}_{\text{prob}}
=\frac{1}{|X|} \sum_{x \in X}
\mathbf{1}\!\left[
D_{\mathrm{KL}}\!\bigl(P_T(\cdot\mid x),P_S(\cdot\mid x)\bigr)\le\delta
\right]
\qquad
\label{MR2Equation3}
\end{equation}

A larger $\text{Loyalty}_{\text{prob}}$ indicates that the student model retains the teacher model’s probability-distribution behavior on a greater portion of the test set.

\subsubsection{MR3 - High Confidence Preservation}
\label{MR3Label}
\paragraph*{\textbf{Description}}  
Given a set of test inputs $X={x_1, x_2, ..., x_n}$, $M_T$, and $M_S$, consider the subset: $\mathcal{X}_{\text{conf}} = \left\{\,x_i \in X \;\middle|\; \max M_T(x_i) \ge \tau \right\},$

where $\tau$ is a high-confidence threshold. We select $\tau$ within the range of $[0.8, 0.9]$, which is commonly adopted in prior work on model confidence and calibration to distinguish reliable predictions from uncertain ones \cite{guo2017calibration, hendrycks2016baseline}. To ensure that the conclusions are not dependent on a specific threshold choice, we evaluate MR3 using multiple confidence levels ($\tau = 0.8, 0.85, 0.9$). For every $x_i \in \mathcal{X}_{\text{conf}}$, the student model should maintain

\begin{equation}
    \arg\max M_S(x_i) \;=\; \arg\max M_T(x_i)
    \quad \land\ \quad
    \max M_S(x_i) \;\ge\; \tau
    \label{MR3Equation1}
\end{equation}

In other words, $M_S$ must predict the same top-1 class as $M_T$ \emph{and} do so with confidence at least $\tau$.

\paragraph*{\textbf{Rationale}} A Top-1 agreement alone may mask cases where $M_S$ is hesitant while $M_T$ is decisive. Our manual investigation (Figure 3(a)) revealed that large confidence gaps (e.g., $|0.99-0.51|=0.48$) often coincide with divergent reasoning, even when predictions align. By enforcing both label and confidence alignment on high-certainty inputs, \textbf{MR3} probes a deeper layer of behavioral fidelity.

\paragraph*{\textbf{Evaluation metric}} We introduce the \emph{High Confidence Agreement Rate} (HCAR) metric to quantify how often this relation holds:

\begin{equation}
    \text{HCAR} = \frac{1}{|\mathcal{X}_{\text{conf}}|} \sum_{x \in \mathcal{X}_{\text{conf}}} \mathbb{I} \left[
\begin{aligned}
& \arg\max T(x) = \arg\max S(x) \\
& \land\ \max S(x) \geq \tau
\end{aligned}
\right]
    \label{MR3Equation2}
\end{equation}

Here \( \mathbb{I}[\cdot] \) denotes the indicator function, which returns 1 if the condition is satisfied and 0 otherwise. A higher HCAR indicates that the student model accurately preserves the teacher’s confident and decisive behavior, whereas a lower value suggests reduced fidelity after knowledge distillation.

\subsubsection{MR4 - Calibration Alignment}
\label{MR4Label}
\paragraph*{\textbf{Description}} Given a set of test inputs $X={x_1, x_2, ..., x_n}$, $M_T$, and $M_S$, we partition the confidence interval $[0,1]$ into $B$ equal-width bins. In our study, we consider multiple bin configurations (e.g., $B \in \{10, 15, 20\}$, which are standard values used in calibration analysis \cite{guo2017calibration, naeini2015obtaining}) to assess the robustness of calibration alignment across different granularities. For each bin $i\in\{1,\dots,B\}$, let 

\begin{equation}
    \text{acc}^{\text{orig}}_i = 
\frac{1}{|X_i|}\sum_{x\in X_i} 1\!\bigl[\arg\max M_T(x)=y(x)\bigr]
    \label{MR5Equation1}
\end{equation}

\begin{equation}
    \text{acc}^{\text{comp}}_i =
\frac{1}{|X_i|}\sum_{x\in X_i} 1\!\bigl[\arg\max M_S(x)=y(x)\bigr]
    \label{MR5Equation2}
\end{equation}

where $X_i$ is the subset of inputs whose $M_T$ confidence $\max M_T(x)$ falls into bin~$i$ and $y(x)$ is the ground-truth label. The student model is expected to preserve calibration if, for every bin $i$, its empirical accuracy closely matches that of the teacher.

\paragraph*{\textbf{Rationale}} Top-1 agreement alone does not indicate whether the student model is \emph{well-calibrated}. Poor calibration, where predicted probabilities do not reflect true likelihoods, can lead to over- or under-confidence, which adversely affects downstream decision-making. Prior studies \cite{shang2024enhancing, ji2024beware} on vision and language models demonstrate that compression techniques such as quantization can significantly distort model calibration, even when accuracy remains unaffected. These findings motivate the design of this MR to assess behavioral fidelity through the lens of calibration alignment. To ensure that our analysis is not sensitive to a specific binning configuration, we evaluate calibration alignment across multiple values of $B$ to observe consistent trends in the relative behavior of teacher and student models.

\paragraph*{\textbf{Evaluation metric}} We propose the \emph{Expected Calibration Alignment} (ECA) metric, defined as the mean absolute difference in bin-wise accuracy to assess this behavioral fidelity:

\begin{equation}
    \mathrm{ECA} \;=\;
    \frac{1}{B}\sum_{i=1}^{B}
    \left|\,
    \text{acc}^{\text{orig}}_i - \text{acc}^{\text{comp}}_i
    \right|
\label{MR5quation3}
\end{equation}

A lower ECA score ($\text{ECA} \in [0, 1]$) indicates that the student model’s calibration behavior closely aligns with that of the teacher model. Conversely, a higher ECA score reflects greater divergence between the two models’ calibration profiles.

\textbf{Violation rate computation:} To compute the violation rate of a metamorphic relation \( \text{MR}_k \), we use the complement of its hold rate. Let \( \text{HoldRate}_{\text{MR}_k} \in [0, 1] \) denote the proportion of test inputs for which \( \text{MR}_k \) holds. Then, the violation rate is defined as:

\begin{equation}
\text{ViolationRate}_{\text{MR}_k} = 1 - \text{HoldRate}_{\text{MR}_k}
\label{MRViolation}
\end{equation}

This formulation enables us to quantify the extent to which a student model deviates from the behavioral invariants defined by each metamorphic relation. Higher violation rates indicate greater behavioral drift from the teacher model.

\textbf{Expected Behaviour: }A student model is considered behavior-preserving if, for every test input, the student model: (i) produces the exact top-1 prediction as the teacher; (ii) retains a probability distribution that closely matches the teacher’s in terms of uncertainty and class ranking; (iii) replicates the teacher’s high-confidence decisions on unambiguous inputs (e.g., $\max M_T(x)!\ge!\tau$); and (iv) maintains aligned calibration curves such that the bin-wise accuracy difference  $\lvert\text{acc}^{\text{orig}}_i-\text{acc}^{\text{comp}}_i\rvert$ remains consistently small.

\subsection{\textit{MetaCompress} Evaluation}
\label{MetaEval}
To evaluate the effectiveness of the proposed \textit{MetaCompress} framework, we perform an extensive empirical study using the models, datasets, and knowledge distillation techniques described in Section \ref{MimicStudyDesign}. This evaluation aims to determine whether \textit{MetaCompress} can uncover behavioral discrepancies between teacher and student models that traditional accuracy-based metrics fail to capture.

\subsubsection{Implementations}
\label{Implement}
\textit{MetaCompress} is implemented in Python on top of the \textit{Compressor}, \textit{\textlarger[1]{A}VATAR}, and \textit{MORPH}. All experiments are conducted on an Ubuntu 22.04 machine equipped with an Intel(R) Xeon(R) W-2265 CPU @ 3.50GHz, 256 GB RAM, and one NVIDIA RTX A6000 48 GB GPU.

\subsubsection{Evaluation Results}
\label{MetaEmpResults}
This section presents the empirical evaluation results of the \textit{MetaCompress} framework to address the second research question, \textbf{RQ2}: \textit{To what extent does \textit{MetaCompress} uncover behavioral discrepancies that may not be captured by the traditional \textit{accuracy} metric?}

We address this research question by assessing the effectiveness of the \textit{MetaCompress} framework by measuring the degree of violation across the four designed metamorphic relations. Each violation indicates a behavioral inconsistency between the teacher and student models when exposed to identical inputs, thus quantifying the loss of behavioral fidelity after knowledge distillation. We conduct this analysis across two language models of code (CodeBERT and GraphCodeBERT), two software analytics tasks (clone detection and vulnerability prediction), and three knowledge distillation approaches (\textit{Compressor}, \textit{AVATAR}, and \textit{MORPH}).

\begin{figure}[htbp]
  \centering
  \includegraphics[width=15cm, height=6cm]{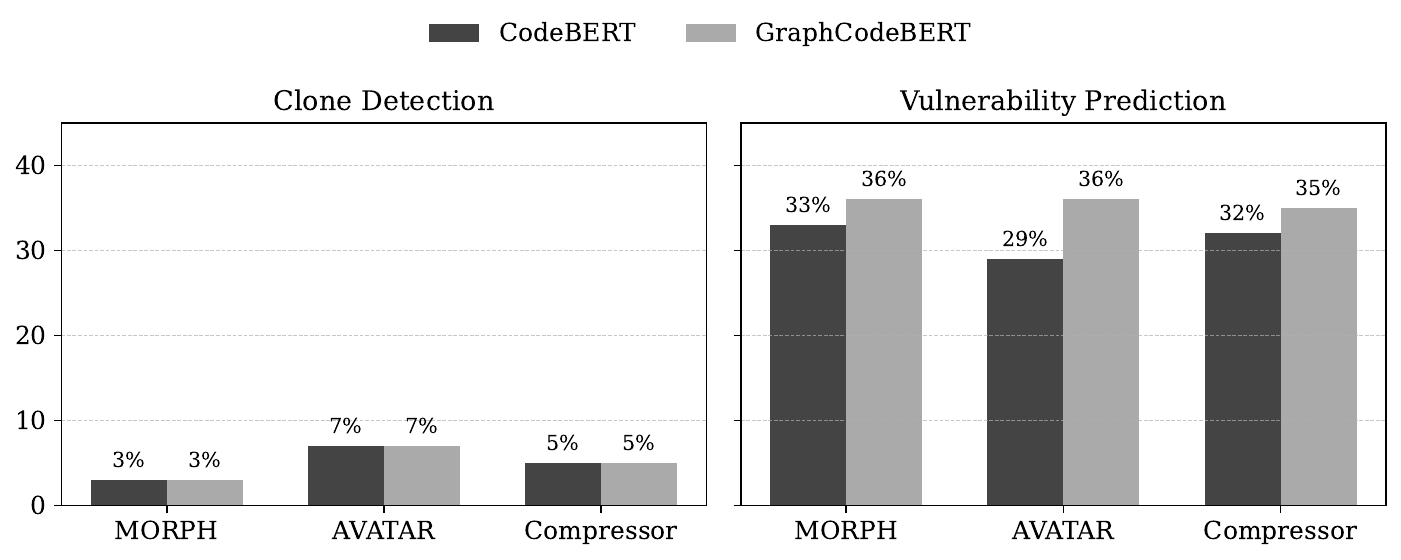}
  \caption{MR1 violation rates for different tasks, models, and knowledge distillation techniques.}
  \label{Mr1Vio}
\end{figure}

\textbf{MR1 violation:} The results for \textbf{MR1} violation shown in Figure \ref{Mr1Vio} reveal that student models exhibit low violation rates for the clone detection task but considerably higher rates for vulnerability prediction. For clone detection, all three distillation techniques maintain strong label agreement with the teacher, with a maximum violation of 7\%, where \textit{MORPH} achieves the lowest (3\%) and \textit{AVATAR} the highest (7\%) inconsistency. In contrast, violation rates rise sharply for vulnerability prediction, ranging from 29\% to 36\%, indicating notable behavioral divergence between teacher and student models. Among the methods, \textit{AVATAR} performs slightly better for \textit{CodeBERT}, while \textit{MORPH} and \textit{Compressor} show comparable fidelity across both models. As shown in Table \ref{AccTab}, although the student models achieve comparable performance on the vulnerability prediction task (with a maximum accuracy difference of approximately 3\%), the \textit{MetaCompress} framework reveals up to 36\% discrepancies in prediction alignment, demonstrating its ability to capture behavioral differences that traditional accuracy-based evaluation overlooks.


\begin{figure}[htbp]
  \centering
  \includegraphics[width=15cm, height=6cm]{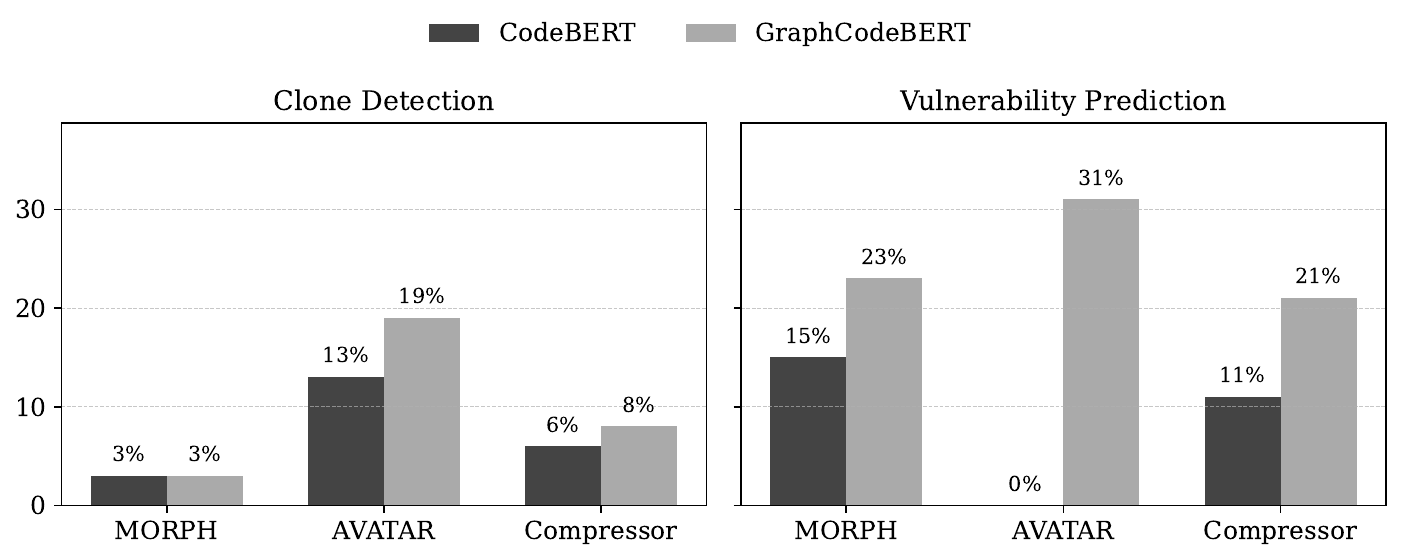}
  \caption{MR2 violation rates for different tasks, models, and knowledge distillation techniques.}
  \label{MR2Violation}
\end{figure}

\textbf{MR2 violation:}  The results for \textbf{MR2} violations shown in Figure \ref{MR2Violation} reveal that the degree of probabilistic divergence between teacher and student models varies across tasks and distillation techniques. For the clone detection task, the violation rates remain relatively low, ranging from 3\% to 13\%, suggesting that the student models generally approximate the teacher’s output distributions. In contrast, the vulnerability prediction task shows substantially greater violations, reaching up to 31\%, reflecting a pronounced mismatch in the probability distributions of the teacher and student models. These findings suggest that, although the student models can match the teacher’s predictions in terms of label accuracy, their underlying probability distributions differ significantly, underscoring \textit{MetaCompress} 's ability to expose fine-grained probabilistic discrepancies beyond traditional accuracy-based evaluation.

\begin{figure}[htbp]
  \centering
  \subfigure[$D_{\mathrm{KL}}(P \parallel Q)$]{\label{KLDPQ}\includegraphics[width=12cm, height=4cm]{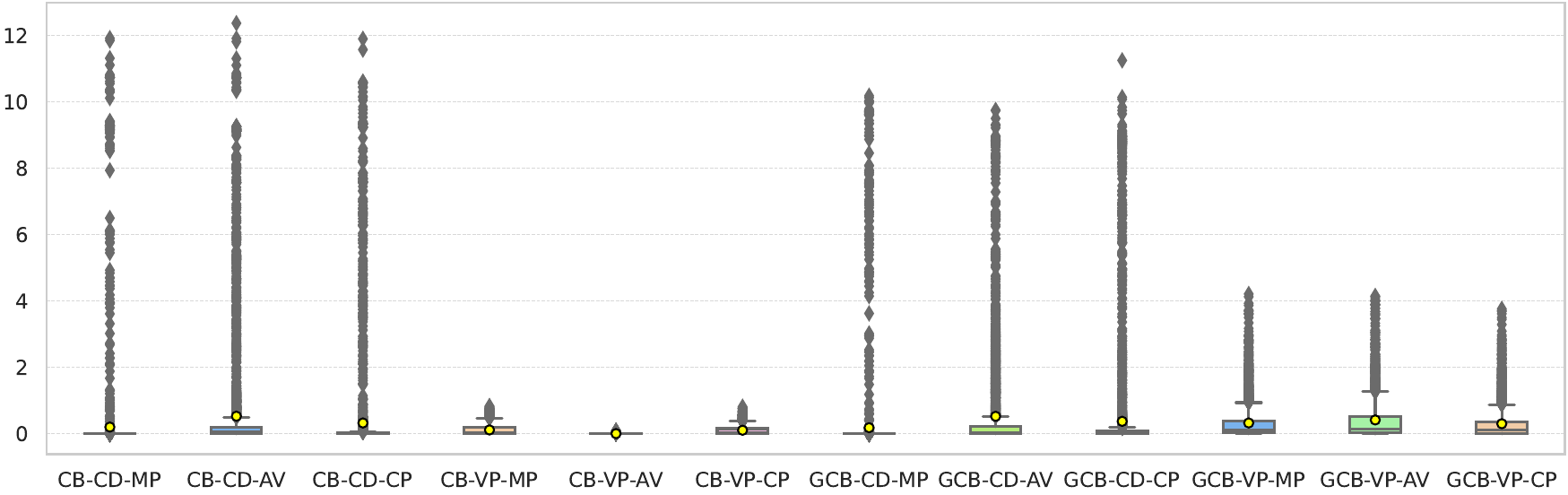}}
  \subfigure[$D_{\mathrm{KL}}(Q \parallel P)$]{\label{KLDQP}\includegraphics[width=12cm, height=4cm]{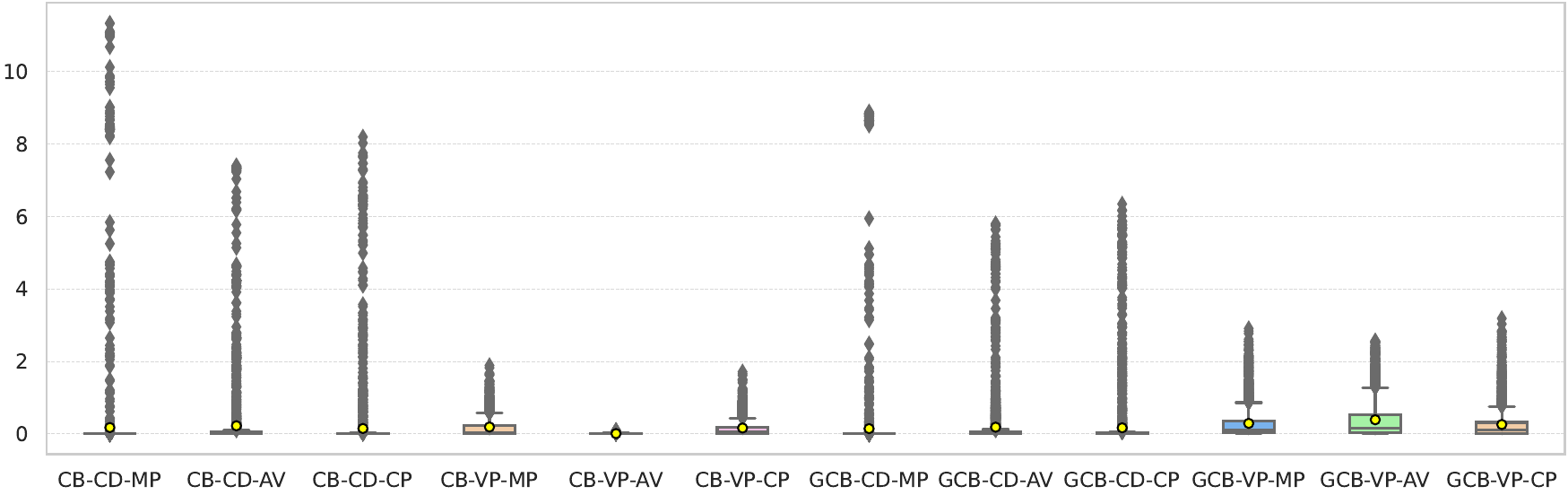}}
  \caption{Boxplots of KL divergence illustrating output distribution shifts between teacher and student models across different tasks and knowledge distillation techniques \protect\footnotemark.}
  \label{MR2Vio}
\end{figure}

To gain a finer insight into behavioral fidelity, we visualize the distributions of MR2 violations using box plots, enabling a head-to-head comparison between teacher and student models in terms of output probability deviation across individual test instances. Given the asymmetry of KL divergence, we report both \( D_{\mathrm{KL}}(P \parallel Q) \) and \( D_{\mathrm{KL}}(Q \parallel P) \), where \( P \) and \( Q \) are the output probability distributions of the teacher and student models. Figure \ref{MR2Vio} presents the per-sample distributions of these values for KD approaches: \textit{Compressor}, \textit{\textlarger[1]{A}VATAR}, and \textit{MORPH}. We observe that for a large proportion of samples, the KL divergence remains low, indicating close alignment between the teacher and student models. However, a notable number of outliers with high divergence suggest that in some cases, the student model assigns significantly different probabilities to the same output classes. These distributions underscore the utility of MR2 for capturing probabilistic distortions that may not be reflected in the \textit{accuracy} metric. Together, the distributions of \( D_{\mathrm{KL}}(P \parallel Q) \) and \( D_{\mathrm{KL}}(Q \parallel P) \) provide a comprehensive understanding of discrepancies in behavioral fidelity at the granularity of individual predictions. These findings demonstrate that \textit{MetaCompress} effectively exposes subtle differences in probabilistic reasoning between teacher and student models that remain hidden under accuracy-based evaluation.

\footnotetext{CB: CodeBERT, GCB: GraphCodeBERT, VP: Vulnerability Prediction, CD: Clone Detection, AV: \textit{\textlarger[1]{A}VATAR}, CP: \textit{Compressor}, MP: \textit{MORPH}, CB-CD-AV: CodeBERT compressed via \textit{\textlarger[1]{A}VATAR} for the Clone Detection task.}


\textbf{MR3 violation:} While MR1 and MR2 expose surface-level inconsistencies, such as label disagreement and output distribution shifts, MR3 uncovers deeper deviations in behavioral fidelity. Table \ref{MR3Tab} illustrates the violations of MR3 observed in student models, as identified by the \textit{MetaCompress}. Across all tasks and student models, we observe substantial violations of MR3. As shown in Table \ref{MR3Tab}, student GraphCodeBERT models often exhibit significantly lower confidence than their teacher counterparts for the vulnerability prediction task, despite having similar accuracy. For example, the GraphCodeBERT model obtained from \textit{AVATAR} violates up to \textbf{62\%} of MR3, indicating a significant discrepancy in high-confidence predictions and overall behavioral fidelity between the teacher and student models that accuracy-based evaluation fails to capture. We further examine the sensitivity of MR3 to different threshold values of $\tau$ and find that variations in $\tau$ had a negligible impact on overall violation rates, confirming the metric's stability.

\begin{table}[htbp]
\centering
\caption{Behavioral fidelity discrepancies in student models revealed through MR3 violations.}
\begin{threeparttable}[t]
\resizebox{\linewidth}{!}{
\begin{tabular}{c|c|lll|lll|lll}
\hline
                                &                                  & \multicolumn{3}{c|}{\textbf{\textit{Compressor}}}                                                                                                                        & \multicolumn{3}{c|}{\textbf{\textit{AVATAR}}}                                                                                                                           & \multicolumn{3}{c}{\textbf{\textit{MORPH}}}                                                                                                                            \\ \cline{3-11} 
\multirow{-2}{*}{\textbf{Task}} & \multirow{-2}{*}{\textbf{Model}} & \multicolumn{1}{l|}{$\tau   = 0.8$}                        & \multicolumn{1}{l|}{$\tau   = 0.85$}                       & $\tau   = 0.9$                        & \multicolumn{1}{l|}{$\tau   = 0.8$}                       & \multicolumn{1}{l|}{$\tau   = 0.85$}                       & $\tau   = 0.9$                        & \multicolumn{1}{l|}{$\tau   = 0.8$}                       & \multicolumn{1}{l|}{$\tau   = 0.85$}                       & $\tau   = 0.9$                        \\ \hline
                                & \textbf{CB}                      & \multicolumn{1}{l|}{0.03}                                  & \multicolumn{1}{l|}{0.03}                                  & 0.03                                  & \multicolumn{1}{l|}{0.06}                                 & \multicolumn{1}{l|}{0.05}                                  & 0.04                                  & \multicolumn{1}{l|}{0.03}                                 & \multicolumn{1}{l|}{0.03}                                  & 0.03                                  \\ \cline{2-11} 
\multirow{-2}{*}{\textbf{CD}}   & \textbf{GCB}                     & \multicolumn{1}{l|}{0.04}                                  & \multicolumn{1}{l|}{0.04}                                  & 0.04                                  & \multicolumn{1}{l|}{0.03}                                 & \multicolumn{1}{l|}{0.03}                                  & 0.03                                  & \multicolumn{1}{l|}{0.03}                                 & \multicolumn{1}{l|}{0.03}                                  & 0.03                                  \\ \hline
                                & \textbf{CB}                      & \multicolumn{1}{l|}{\cellcolor[HTML]{FFFFFF}\textbf{--}}   & \multicolumn{1}{l|}{\cellcolor[HTML]{FFFFFF}\textbf{--}}   & \cellcolor[HTML]{FFFFFF}\textbf{--}   & \multicolumn{1}{l|}{\cellcolor[HTML]{FFFFFF}\textbf{--}}  & \multicolumn{1}{l|}{\cellcolor[HTML]{FFFFFF}\textbf{--}}   & \cellcolor[HTML]{FFFFFF}\textbf{--}   & \multicolumn{1}{l|}{\cellcolor[HTML]{FFFFFF}\textbf{--}}  & \multicolumn{1}{l|}{\cellcolor[HTML]{FFFFFF}\textbf{--}}   & \cellcolor[HTML]{FFFFFF}\textbf{--}   \\ \cline{2-11} 
\multirow{-2}{*}{\textbf{VP}}   & \textbf{GCB}                     & \multicolumn{1}{l|}{\cellcolor[HTML]{C0C0C0}\textbf{0.44}} & \multicolumn{1}{l|}{\cellcolor[HTML]{C0C0C0}\textbf{0.45}} & \cellcolor[HTML]{C0C0C0}\textbf{0.43} & \multicolumn{1}{l|}{\cellcolor[HTML]{C0C0C0}\textbf{0.6}} & \multicolumn{1}{l|}{\cellcolor[HTML]{C0C0C0}\textbf{0.61}} & \cellcolor[HTML]{C0C0C0}\textbf{0.62} & \multicolumn{1}{l|}{\cellcolor[HTML]{C0C0C0}\textbf{0.5}} & \multicolumn{1}{l|}{\cellcolor[HTML]{C0C0C0}\textbf{0.51}} & \cellcolor[HTML]{C0C0C0}\textbf{0.49} \\ \hline
\end{tabular}
}

\begin{tablenotes} \small
  \item[] \parbox[t]{0.75\linewidth}{
    *Cells highlighted with a dashed line indicate cases where the teacher models do not produce any output with a high confidence score for the given task.
  }
\end{tablenotes}
\end{threeparttable}
\label{MR3Tab}
\end{table}

\textbf{MR4 violation:} Table \ref{MR4Tab} presents the calibration deviation between teacher and student models across three knowledge distillation techniques for varying bin sizes ($B=10, 15, 20$). The results reveal that the calibration alignment varies notably across tasks and distillation strategies. For the clone detection task, all student models maintain near-perfect calibration consistency with their teachers, with deviation values below 0.04, indicating that the students effectively preserve the teachers' confidence distributions. In contrast, the vulnerability prediction task exhibits substantially higher deviations marked by dark gray cells, up to 0.14 for CodeBERT and 0.11 for GraphCodeBERT, demonstrating that the student models' probabilistic calibration diverges considerably from that of the teachers. These results highlight that although student models may achieve comparable accuracy, their confidence calibration can differ sharply, underscoring \textit{MetaCompress} 's ability to capture nuanced behavioral discrepancies beyond what traditional evaluation metrics can reveal. We further examine the sensitivity of \textbf{MR4} to different bin sizes and observe that variations in $B$ had a negligible effect on overall violation rates, confirming the metric’s stability.

\begin{table}[htbp]
\centering
\caption{Behavioral fidelity discrepancies in student models revealed through MR4 violations.}
\begin{threeparttable}[t]
\resizebox{\linewidth}{!}{
\begin{tabular}{c|c|lll|lll|lll}
\hline
                                &                                  & \multicolumn{3}{c|}{\textbf{\textit{Compressor}}}                                                                                                                           & \multicolumn{3}{c|}{\textbf{\textit{AVATAR}}}                                                                                                                               & \multicolumn{3}{c}{\textbf{\textit{MORPH}}}                                                                                                                                \\ \cline{3-11} 
\multirow{-2}{*}{\textbf{Task}} & \multirow{-2}{*}{\textbf{Model}} & \multicolumn{1}{l|}{B =   10}                               & \multicolumn{1}{l|}{B =   15}                               & B =   20                               & \multicolumn{1}{l|}{B =   10}                               & \multicolumn{1}{l|}{B =   15}                               & B =   20                               & \multicolumn{1}{l|}{B =   10}                               & \multicolumn{1}{l|}{B =   15}                               & B =   20                               \\ \hline
                                & \textbf{CB}                      & \multicolumn{1}{l|}{0.017}                                  & \multicolumn{1}{l|}{0.017}                                  & 0.017                                  & \multicolumn{1}{l|}{0.035}                                  & \multicolumn{1}{l|}{0.035}                                  & 0.035                                  & \multicolumn{1}{l|}{0.002}                                  & \multicolumn{1}{l|}{0.002}                                  & 0.002                                  \\ \cline{2-11} 
\multirow{-2}{*}{\textbf{CD}}   & \textbf{GCB}                     & \multicolumn{1}{l|}{0.026}                                  & \multicolumn{1}{l|}{0.026}                                  & 0.026                                  & \multicolumn{1}{l|}{0.063}                                  & \multicolumn{1}{l|}{0.063}                                  & 0.063                                  & \multicolumn{1}{l|}{0.003}                                  & \multicolumn{1}{l|}{0.003}                                  & 0.003                                  \\ \hline
                                & \textbf{CB}                      & \multicolumn{1}{l|}{\cellcolor[HTML]{C0C0C0}\textbf{0.136}} & \multicolumn{1}{l|}{\cellcolor[HTML]{C0C0C0}\textbf{0.142}} & \cellcolor[HTML]{C0C0C0}\textbf{0.143} & \multicolumn{1}{l|}{0.016}                                  & \multicolumn{1}{l|}{0.016}                                  & 0.017                                  & \multicolumn{1}{l|}{\cellcolor[HTML]{C0C0C0}\textbf{0.133}} & \multicolumn{1}{l|}{\cellcolor[HTML]{C0C0C0}\textbf{0.138}} & \cellcolor[HTML]{C0C0C0}\textbf{0.139} \\ \cline{2-11} 
\multirow{-2}{*}{\textbf{VP}}   & \textbf{GCB}                     & \multicolumn{1}{l|}{0.092}                                  & \multicolumn{1}{l|}{0.092}                                  & 0.092                                  & \multicolumn{1}{l|}{\cellcolor[HTML]{C0C0C0}\textbf{0.112}} & \multicolumn{1}{l|}{\cellcolor[HTML]{C0C0C0}\textbf{0.112}} & \cellcolor[HTML]{C0C0C0}\textbf{0.112} & \multicolumn{1}{l|}{0.094}                                  & \multicolumn{1}{l|}{0.096}                                  & 0.098                                  \\ \hline
\end{tabular}
}

\begin{tablenotes} \small
  \item[] \parbox[t]{0.75\linewidth}{
    *Cells highlighted with a dashed line indicate cases where the teacher models do not produce any output with a high confidence score for the given task.
  }
\end{tablenotes}

\end{threeparttable}
\label{MR4Tab}
\end{table}

\section*{Robustness Analysis of \textit{MetaCompress}}
To assess the robustness of \textit{MetaCompress} as an evaluation framework, we conduct an ablation study to determine whether its conclusions remain consistent when the inputs to both the teacher and student models are transformed for the vulnerability prediction task. Specifically, we recompute MR1–MR4 violations using transformed inputs and compare the resulting behavioral fidelity patterns with those obtained from the original inputs. To obtain transformed inputs, we use metamorphic code\footnote{Unlike adversarial examples, which are generated through iterative identifier modifications guided by queries to the target model, metamorphic code is obtained through a one-time transformation of identifier names that preserves program semantics.} generated following Panichella et al. \cite{panichella2025metamorphic}, which applies semantics-preserving transformations such as method and parameter renaming. Identifiers are replaced with meaningful synonyms (e.g., \textit{args} to \textit{arguments}) using \textit{ChatGPT-3.5} to ensure naturalness. The generated code is validated for syntactic correctness using \textit{tree-sitter}, with a subset manually inspected for semantic equivalence.

The results presented in Table~\ref{MetaCodeAcc} indicate that the predictive performance of compressed models is comparable to that of uncompressed models when evaluated on metamorphically transformed inputs. The differences in accuracy between the uncompressed and compressed models generated by \textit{Compressor}, \textit{AVATAR}, and \textit{MORPH} are minimal for both CodeBERT and GraphCodeBERT. These findings indicate that, according to conventional accuracy-based evaluation, compressed models maintain task-level performance following input transformations.

\begin{table}[htbp]
\centering
\caption{Accuracy (\%) comparison between teacher (uncompressed) and student (compressed) models on metamorphic transformed inputs for the vulnerability prediction task.}
\label{MetaCodeAcc}
\begin{tabular}{l c c c c}
\toprule
\textbf{Base Model} & \textbf{Teacher} & \textbf{Compressor} & \textbf{AVATAR} & \textbf{MORPH} \\
\midrule
CodeBERT      & 0.6 & 0.59 & 0.59 & 0.58 \\
GraphCodeBERT & 0.61 & 0.6 & 0.58 & 0.59 \\
\bottomrule
\end{tabular}
\end{table}

\begin{figure}[htbp]
  \centering
  \subfigure[MR1 Violation]{\label{MR1MetaCode}\includegraphics[width=7cm, height=5cm]{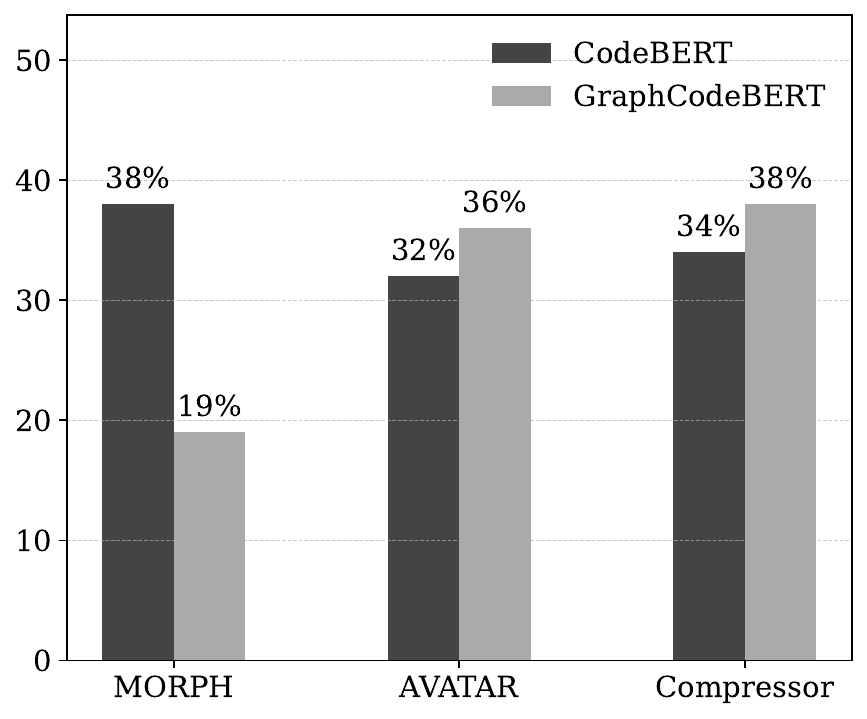}}
  \subfigure[MR2 Violation]{\label{MR2MetaCode}\includegraphics[width=7cm, height=5cm]{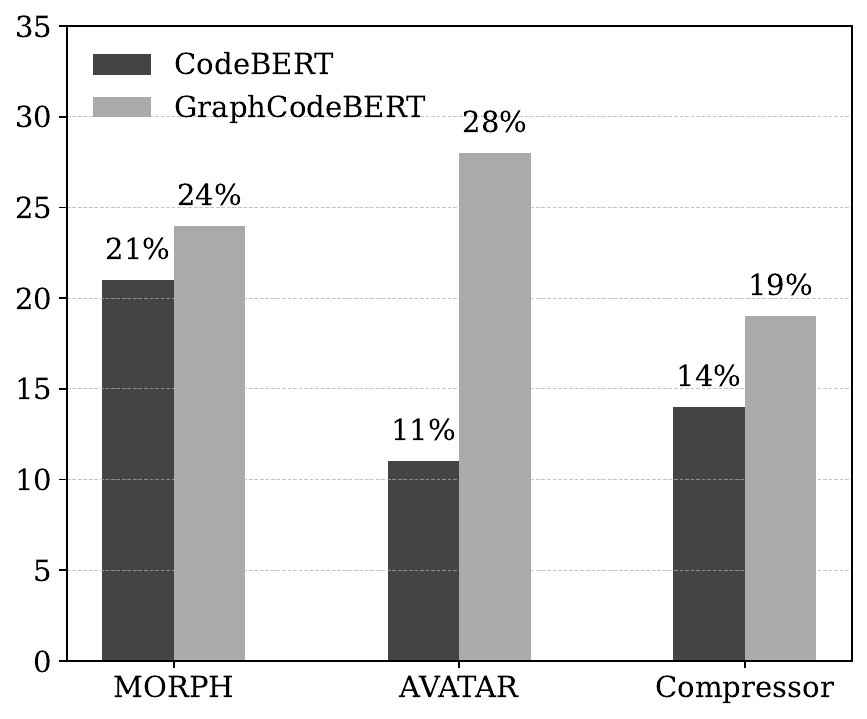}}
  \caption{MR1 and MR2 violation rates for CodeBERT and GraphCodeBERT student models on transformed inputs in the vulnerability prediction task across different knowledge distillation approaches.}
  \label{MR1MR2MetaCode}
\end{figure}

To further examine behavioral fidelity divergence on the transformed inputs, we recompute MR1–MR4 violations, as presented in the Figure \ref{MR1MR2MetaCode} and Table \ref{MR3MR4MetaCode}. The results for MR1 and MR2 on transformed inputs further highlight the behavioral fidelity gap between teacher and student models. For MR1, which assesses label agreement, student models demonstrate substantial violation rates across all knowledge distillation methods. In the vulnerability prediction task, CodeBERT-based student models exhibit violation rates of 32\%-38\%. GraphCodeBERT-based models show similar or greater variability, with violation rates ranging from 19\% to 38\%. The highest violation rate occurs with the \textit{Compressor} approach in GraphCodeBERT (38\%), indicating considerable divergence in predicted labels following input transformation. A comparable trend is evident for MR2, which evaluates probability distribution alignment. Although MR2 violation rates are generally lower than those for MR1, they remain significant across all configurations. CodeBERT-based models exhibit violation rates of 11\%-21\%, while GraphCodeBERT-based models show greater variability, with rates of 19\%-28\%. The \textit{AVATAR} approach yields the highest MR2 violation (28\%) for GraphCodeBERT, indicating substantial discrepancies in the confidence distributions between the teacher and student models. Overall, these results demonstrate that, even when accuracy is comparable, student models consistently fail to maintain both class-level and distribution-level behavioral fidelity under transformed inputs.

The results presented in Table~\ref{MR3MR4MetaCode} indicate substantial discrepancies in behavioral fidelity that are not identified by accuracy-based evaluation. For MR3, GraphCodeBERT consistently shows high violation rates across all compression techniques and threshold settings, reaching up to 0.52. This suggests a significant divergence in prediction confidence when subjected to transformed inputs. In comparison, MR4 violations are lower but remain notable, with values up to 0.15 for CodeBERT and 0.12 for GraphCodeBERT across various bin configurations. Overall, these findings demonstrate that, despite achieving similar predictive accuracy, compressed models do not maintain the confidence consistency and calibration behavior exhibited by their teacher models. Notably, even when both teacher and student models are evaluated on transformed inputs and demonstrate comparable accuracy, \textit{MetaCompress} reveals behavioral fidelity differences. This observation suggests its usefulness as an evaluation framework that extends beyond traditional accuracy-based metrics.

\begin{table}[htbp]
\caption{Behavioral fidelity discrepancies in student models revealed through MR3 and MR4 violations on the transformed inputs.}
\resizebox{\linewidth}{!}{
\begin{tabular}{c|c|lllllllll|lllllllll}
\hline
                                &                                  & \multicolumn{9}{c|}{\textbf{MR3 Violation}}                                                                                                                                                                                                                                                                                                                                                                                                                                                                                                   & \multicolumn{9}{c}{\textbf{MR4 Violation}}                                                                                                                                                                                                                                                                                                                                                                                                            \\ \cline{3-20} 
                                &                                  & \multicolumn{3}{c|}{\textbf{\textit{Compressor}}}                                                                                                                                             & \multicolumn{3}{c|}{\textbf{\textit{AVATAR}}}                                                                                                                                                 & \multicolumn{3}{c|}{\textbf{\textit{MORPH}}}                                                                                                                             & \multicolumn{3}{c|}{\textbf{\textit{Compressor}}}                                                                                                                                             & \multicolumn{3}{c|}{\textbf{\textit{AVATAR}}}                                                                                                                                                 & \multicolumn{3}{c}{\textbf{\textit{MORPH}}}                                      \\ \cline{3-20} 
\multirow{-3}{*}{\textbf{Task}} & \multirow{-3}{*}{\textbf{Model}} & \multicolumn{1}{l|}{$\tau   = 0.8$}                        & \multicolumn{1}{l|}{$\tau   = 0.85$}                       & \multicolumn{1}{l|}{$\tau   = 0.9$}                        & \multicolumn{1}{l|}{$\tau   = 0.8$}                        & \multicolumn{1}{l|}{$\tau   = 0.85$}                       & \multicolumn{1}{l|}{$\tau   = 0.9$}                        & \multicolumn{1}{l|}{$\tau   = 0.8$}                        & \multicolumn{1}{l|}{$\tau   = 0.85$}                       & $\tau   = 0.9$                        & \multicolumn{1}{l|}{B =   10}                              & \multicolumn{1}{l|}{B =   15}                              & \multicolumn{1}{l|}{B =   20}                              & \multicolumn{1}{l|}{B =   10}                              & \multicolumn{1}{l|}{B =   15}                              & \multicolumn{1}{l|}{B =   20}                              & \multicolumn{1}{l|}{B =   10} & \multicolumn{1}{l|}{B =   15} & B =   20 \\ \hline
                                & \textbf{CB}                      & \multicolumn{1}{l|}{--}                                    & \multicolumn{1}{l|}{--}                                    & \multicolumn{1}{l|}{--}                                    & \multicolumn{1}{l|}{--}                                    & \multicolumn{1}{l|}{--}                                    & \multicolumn{1}{l|}{--}                                    & \multicolumn{1}{l|}{--}                                    & \multicolumn{1}{l|}{--}                                    & --                                    & \multicolumn{1}{l|}{\cellcolor[HTML]{C0C0C0}\textbf{0.14}} & \multicolumn{1}{l|}{\cellcolor[HTML]{C0C0C0}\textbf{0.15}} & \multicolumn{1}{l|}{\cellcolor[HTML]{C0C0C0}\textbf{0.15}} & \multicolumn{1}{l|}{0.02}                                  & \multicolumn{1}{l|}{0.02}                                  & \multicolumn{1}{l|}{0.02}                                  & \multicolumn{1}{l|}{0.13}     & \multicolumn{1}{l|}{0.14}     & 0.14     \\ \cline{2-20} 
\multirow{-2}{*}{\textbf{VP}}   & \textbf{GCB}                     & \multicolumn{1}{l|}{\cellcolor[HTML]{C0C0C0}\textbf{0.52}} & \multicolumn{1}{l|}{\cellcolor[HTML]{C0C0C0}\textbf{0.51}} & \multicolumn{1}{l|}{\cellcolor[HTML]{C0C0C0}\textbf{0.48}} & \multicolumn{1}{l|}{\cellcolor[HTML]{C0C0C0}\textbf{0.38}} & \multicolumn{1}{l|}{\cellcolor[HTML]{C0C0C0}\textbf{0.35}} & \multicolumn{1}{l|}{\cellcolor[HTML]{C0C0C0}\textbf{0.36}} & \multicolumn{1}{l|}{\cellcolor[HTML]{C0C0C0}\textbf{0.51}} & \multicolumn{1}{l|}{\cellcolor[HTML]{C0C0C0}\textbf{0.49}} & \cellcolor[HTML]{C0C0C0}\textbf{0.44} & \multicolumn{1}{l|}{0.09}                                  & \multicolumn{1}{l|}{0.09}                                  & \multicolumn{1}{l|}{0.09}                                  & \multicolumn{1}{l|}{\cellcolor[HTML]{C0C0C0}\textbf{0.12}} & \multicolumn{1}{l|}{\cellcolor[HTML]{C0C0C0}\textbf{0.12}} & \multicolumn{1}{l|}{\cellcolor[HTML]{C0C0C0}\textbf{0.12}} & \multicolumn{1}{l|}{0.09}     & \multicolumn{1}{l|}{0.1}      & 0.1      \\ \hline
\end{tabular}
}
\label{MR3MR4MetaCode}
\end{table}

\begin{center}
\begin{tcolorbox}[
    enhanced,
    attach boxed title to top left={yshift=-3mm,yshifttext=-1mm}, 
    colback=mycolor_box,                 
    colframe=black,                
    colbacktitle= mycolor_title,            
    coltitle=black,                
    title=Answer to RQ2,            
    fonttitle=\bfseries,           
    boxed title style={size=small},
]

\textit{MetaCompress} provides a comprehensive evaluation of behavioral fidelity in language models for code, revealing substantial discrepancies that remain hidden under accuracy-based evaluation. Even when the accuracy difference between teacher and student models is within \textbf{3\%}, the framework identifies up to \textbf{62\%} behavioral fidelity violations across the four metamorphic relations. By analyzing model behavior from label agreement (\textbf{MR1}) to probabilistic divergence (\textbf{MR2}), confidence consistency (\textbf{MR3}), and calibration alignment (\textbf{MR4}), \textit{MetaCompress} captures discrepancies across multiple behavioral dimensions. These findings confirm that \textit{MetaCompress} effectively uncovers behavioral differences between teacher and student models that conventional accuracy metrics fail to reveal. Furthermore, the ablation study on transformed inputs demonstrates that these observations remain consistent under input variation, providing additional evidence of the robustness of the \textit{MetaCompress} framework.


\end{tcolorbox}
\label{RQ2Result}
\end{center}

\section{Discussion}
\label{Disc}
\subsection{Behavioral Fidelity and Knowledge Gap in Knowledge Distillation}

Ba et al. \cite{ba2014deep} stated that "\textit{If the student model learns to mimic the teacher model perfectly, it makes exactly the same predictions and mistakes as the teacher model.}" However, achieving such mimicry is extremely challenging when there exists a substantial gap in network capacity \cite{wang2021knowledge}. For instance, the knowledge distillation technique \textit{Compressor} reduces the size of the \textit{CodeBERT} model from 476 MB to only 3 MB. Consequently, although the student models produced by \textit{Compressor}, \textit{AVATAR}, and \textit{MORPH} achieve accuracy comparable to the teacher, their performance deteriorates considerably under adversarial attacks. Even though \textit{MORPH} attempts to distill knowledge more robustly by incorporating metamorphic attacks through metamorphic code transformations, the resulting student models still exhibit notable adversarial vulnerabilities. These observations indicate that the knowledge distillation strategies employed by \textit{Compressor}, \textit{AVATAR}, and \textit{MORPH} do not enable the student models to deeply mimic the teacher’s behavior, leaving a substantial \textit{knowledge gap} in behavioral fidelity between the teacher and the student models.

One might argue that a student model performs comparably to its teacher after knowledge distillation. Then, \textit{why should we be concerned about the \textit{knowledge gap} in behavioral fidelity}? We address this question through two key arguments.

\begin{enumerate}
    \item If the knowledge gap can be reduced by enabling the student model to more deeply mimic the teacher model, then the student should exhibit comparable robustness to its teacher under adversarial conditions. This would make the student model suitable for deployment in real-world software systems where safety and reliability are critical.

    \item  Wang et al. \cite{wang2025empirical} empirically demonstrated that the same student model distilled from a teacher performs significantly better than when trained using the standard fine-tuning approach on the same task. Since the teacher model typically possesses much greater network capacity, it can learn more complex decision boundaries and have greater robustness against adversarial attacks. Minimizing the \textit{knowledge gap}, therefore, enables the distilled student model to better approximate these boundaries during distillation, potentially enhancing its resilience to adversarial attacks.
\end{enumerate}

The empirical findings from our \textit{MetaCompress} framework substantiate these arguments. Although the accuracy difference between teacher and student models remains within 3\%, \textit{MetaCompress} reveals violations of up to 62\% across the four metamorphic relations, indicating substantial discrepancies in behavioral fidelity. For instance, the high MR1 and MR2 violation rates demonstrate that the student models often diverge from the teacher in both label agreement and probability distribution, even when their predictions appear similar at the aggregate level. Likewise, MR3 and MR4 violations expose inconsistencies in confidence and calibration alignment, underscoring that accuracy alone cannot fully capture behavioral divergence. These results confirm that reducing the \textit{knowledge gap} in behavioral fidelity is essential for developing distilled models that preserve the robustness and decision consistency of their teacher counterparts. These insights open promising avenues for advancing knowledge distillation methods that achieve greater alignment between student and teacher models in terms of predictive behavior, internal representations, generalizability, and robustness.

\subsection{Practical Consideration and Deployment Cost}

From a practical standpoint, evaluating the behavioral fidelity does not impose significant overhead. \textit{MetaCompress} is designed exclusively as an evaluation framework to determine whether a distilled student model deeply mimics its teacher prior to deployment. In standard knowledge distillation workflows, both the teacher and student models are evaluated on held-out test data to assess predictive performance, which requires forward passes through both models. \textit{MetaCompress} leverages the probability distributions and related outputs generated during this routine evaluation as byproducts, without requiring additional model executions. Therefore, even for teams that frequently conduct model distillation, \textit{MetaCompress} introduces only negligible computational overhead beyond lightweight post-processing. This approach is both practical and scalable for diagnosing behavioral fidelity gaps and ensuring that compressed student models are robust and reliable for real-world deployment.

\subsection{Actionable Insights for Practitioners and Researchers}
Although \textit{MetaCompress} reports violation rates as diagnostic measures of behavioral fidelity divergence, these values should not be interpreted in isolation. Instead, they serve as actionable indicators, grounded in the empirical observations of this study, such as substantial robustness gaps and MR violations. These indicators inform decision-making throughout the model development and deployment pipeline.

\subsubsection{Insights for Practitioners}

\begin{enumerate}
    \item The appropriate response to an observed metamorphic relation (MR) violation rate depends on the intended deployment context and the acceptable level of risk. For example, a student model with a low violation rate (e.g., below 10\%) may be considered sufficiently stable for use in non-critical applications, such as software clone detection. In contrast, for security-critical tasks (e.g., vulnerability detection) or adversarially exposed environments, practitioners should exercise greater caution and avoid deploying highly compact models. Therefore, it is recommended that practitioners integrate violation-rate checks into continuous integration and deployment pipelines to serve as a gating criterion for model release. Models exhibiting low violation rates may be suitable for deployment in non-critical applications. In contrast, models designed for security-sensitive or adversarial environments should adhere to more stringent thresholds or be supplemented with additional safeguards, such as adversarial defenses or post-deployment monitoring, prior to release.

    \item The acceptable metamorphic relation (MR) violation rate also depends on both the model and the application context. For instance, Figure \ref{Mr1Vio} demonstrates that student models exhibit low MR1 violation rates (up to 7\%) for the clone detection task, but substantially higher rates (up to 36\%) for vulnerability prediction. A similar pattern is observed for MR2 violations (Figure \ref{MR2Vio}), reaching up to 31\% in the vulnerability prediction task. As presented in Table \ref{MR3Tab}, compressed GraphCodeBERT models derived from AVATAR violate up to 62\% of MR3 in the vulnerability prediction task. In contrast, the highest MR4 violation (up to 0.142 as shown in Table \ref{MR4Tab}) occurs in the clone detection task. These findings indicate that practitioners should interpret MR violation rates in the context of both the specific task and the model configuration, and make deployment decisions by weighing behavioral divergence against the application's risk tolerance.

    \item The metamorphic relation violation rates indicate substantial behavioral divergence between teacher and student models within a specific application context, as shown in Figures \ref{Mr1Vio}, \ref{MR2Vio}, and Tables \ref{MR3Tab}, \ref{MR4Tab}. Therefore, practitioners should consider revisiting the selected student architecture or adopting a knowledge distillation strategy informed by the MR violations. For instance, in the clone detection task, the CodeBERT and GraphCodeBERT models derived from \textit{AVATAR} exhibit higher MR1 violation rates (7\%) than those from \textit{Compressor} (3\%) and \textit{MORPH} (5\%). A similar trend is observed for MR2 violations: models obtained via the \textit{AVATAR} knowledge distillation approach exhibit higher MR violations (13\% for CodeBERT and 19\% for GraphCodeBERT) than those from \textit{Compressor} and \textit{MORPH}. Conversely, in the vulnerability detection task, models produced by \textit{AVATAR} demonstrate lower MR3 violations than those generated by \textit{Compressor} and \textit{MORPH}. These findings indicate that \textit{MetaCompress} can assist practitioners in selecting suitable compression techniques for specific models and tasks, thereby ensuring that knowledge-distilled models more accurately mimic the learned representations and decision boundaries of their teacher models.

\end{enumerate}

\subsubsection{Insights for Researchers}

\begin{enumerate}
    \item From a research perspective, MR violation rates offer a quantitative measure for assessing knowledge transfer quality that extends beyond traditional accuracy metrics. As shown in our experimental results, student models often achieve comparable accuracy to teacher models while exhibiting significant MR violations and reduced adversarial robustness, indicating incomplete transfer of decision boundaries. For example, in Figure~\ref{MotivatingExample}, the teacher assigns a confidence score of 0.99, while the student assigns 0.51 for the same instance (e.g., idx: 21383), and this discrepancy is captured by MR2. These observations suggest that relying solely on soft targets for knowledge distillation may not adequately transfer the teacher’s decision boundaries and internal representations. These observations suggest that reducing MR violations requires improving the transfer of knowledge quality from the teacher to the student beyond final predictions. To this end, researchers may incorporate intermediate-layer alignment, such as matching hidden representations or attention distributions \cite{jiao2020tinybert, wang2020minilm, wang2021minilmv2, sun2020mobilebert}, designing distribution-level alignment losses \cite{park2019relational}, and applying robustness-aware distillation objectives \cite{goldblum2020adversarially, zi2021revisiting} that better preserve the teacher model’s confidence and calibration behavior.

    \item Our empirical findings indicate that robustness degradation persists across compressed models trained with various knowledge distillation techniques (e.g., \textit{AVATAR} and \textit{MORPH} in Figures \ref{ASRCompare} and \ref{ASRCODACompare}). This persistence suggests that a single-teacher distillation process may not adequately capture the diverse decision regions within the input space. This limitation aligns with the findings of Yang et al. \cite{yang2025one}, who demonstrated that Mixture of Experts (MoE) \cite{jacobs1991adaptive} frameworks improve software vulnerability detection compared to single-model approaches. Based on these insights, future research should investigate multi-teacher knowledge distillation leveraging MoE architectures to aggregate complementary knowledge from multiple specialized teachers \cite{awal2026moekd}.

    \item Finally, the experimental results indicate that current knowledge distillation approaches primarily focus on aligning final-layer outputs, yet significant behavioral discrepancies persist, as evidenced by high MR violation rates and adversarial vulnerability gaps. These findings imply that essential structural and representational information is lost during the distillation process, and that soft knowledge alignment alone does not adequately preserve behavioral fidelity. Future research should explore hybrid distillation strategies that integrate output-level supervision with intermediate-layer alignment and behavioral constraints designed in our \textit{MetaCompress} framework, aiming to minimize MR violations and enhance behavioral fidelity between teacher and student models.

\end{enumerate}

In summary, the insights above suggest that behavioral fidelity divergence and metamorphic relation violation rates should be interpreted as decision-support metrics rather than pass–fail criteria. By making decision-boundary instability observable and measurable, \textit{MetaCompress} enables developers to make informed follow-up decisions—such as pursuing architectural refinements or accepting a model based on application-specific risk tolerance—and encourages researchers in designing more robust and reliable knowledge distillation techniques.

\section{Threats to Validity}
\label{THV}
This section provides a brief overview of the construct and the internal and external threats associated with our study.

\subsection{Internal Validity}
\label{IntThreat}
A potential threat to internal validity may arise if differences in performance between the student and teacher models are influenced by inconsistent training procedures, fine-tuning configurations, or data handling, rather than the knowledge distillation approaches themselves. To mitigate this threat, we applied uniform training protocols and hyperparameter settings across all models, both student and teacher, following the configurations reported in prior knowledge distillation studies \cite{shi2022compressing, shi2024greening, panichella2025metamorphic}. Additionally, we used identical data splits for fine-tuning and evaluation to ensure consistency and fairness in comparison. Maintaining such a consistent experimental setup enhances the internal validity of our findings by ensuring that the observed effects are due to the student model’s inability to deeply mimic the teacher model’s behavior, rather than differences in hyperparameters or data partitions. Additionally, the experimental results may be influenced by potential errors in implementing the training pipelines for both student and teacher models. To mitigate this threat, we adopted the standardized training procedures provided by the \textit{CodeXGLUE} benchmark for teacher model fine-tuning and evaluation. For knowledge distillation, we utilized publicly available, off-the-shelf student models obtained from \textit{Compressor}\footnote{https://github.com/soarsmu/Compressor}, \textit{AVATAR}\footnote{https://github.com/soarsmu/Avatar}, and \textit{MORPH}\footnote{https://github.com/apanichella/morph-distillation/tree/main}.

\subsection{Construct Validity}
\label{ConsThreat}
A potential threat to construct validity lies in the assumption that the four metamorphic relations accurately capture behavioral fidelity. To mitigate this, the relations were not defined arbitrarily; instead, they were grounded in empirical observations of how student models diverge from their teacher counterparts on identical inputs (Figure \ref{MotivatingExample}). These observed patterns informed the design of each relation, ensuring that they capture behaviorally meaningful differences between student and teacher models. A potential threat to construct validity arises from framing behavioral fidelity evaluation as an oracle problem, which assumes the absence of ground-truth labels despite the availability of test inputs. To mitigate this, we argue that the outputs of the student and the teacher models are not known until execution, much like the behavior of a software system is observed only at runtime. This reinforces the validity of framing behavioral fidelity evaluation as an oracle problem. Another potential threat to construct validity arises from the inherent randomness in model behavior, which can lead to misleading conclusions about behavioral differences. To address this issue, we fixed the random seeds and conducted head-to-head comparisons between each student model and its teacher counterpart. The comparisons used identical inputs across all four metamorphic relations to ensure that any observed differences were due to the student model’s inability to deeply mimic the teacher model’s behavior rather than random variation.

\subsection{External Validity}
\label{ExtThreat}
A potential external threat is the generalizability of \textit{MetaCompress} beyond program understanding to tasks such as code summarization. This study focuses exclusively on knowledge distillation-based approaches, including \textit{Compressor}, \textit{AVATAR}, and \textit{MORPH}, which are designed for code-language models and operate on the teacher model's logits. Since these methods compute the loss between teacher and student logits for knowledge distillation, they are not directly applicable to generation tasks, where it is uncommon to define a loss function over generated sequences that typically differ in length between the teacher and the student models. This limitation reflects an inherent constraint of the state-of-the-art knowledge distillation approaches proposed for code language models rather than a limitation of the proposed framework itself. Despite this, \textit{MetaCompress} is designed to be architecture-agnostic and operates exclusively on model outputs, which enables generalization across various models and tasks without requiring access to internal representations. To enhance external validity, this study evaluates multiple widely adopted models (e.g., CodeBERT and GraphCodeBERT), diverse software analytics tasks (e.g., clone detection and vulnerability prediction), and representative state-of-the-art knowledge distillation approaches (e.g., \textit{Compressor}, \textit{AVATAR}, and \textit{MORPH}). The consistent observation of behavioral fidelity gaps across these settings indicates that the findings are not limited to a particular model, task, or distillation method. Nevertheless, the selection of models, tasks, and datasets may still limit broader generalizability. Future work may extend this evaluation to additional model architectures, tasks, and real-world deployment scenarios. To facilitate such efforts, we have thoroughly documented our framework design and evaluation setup and publicly released all code and scripts used in our experiments.

\subsection{Conclusion Validity}
\label{ConVal}
Our study indicates that student models produced via knowledge distillation tend to exhibit reduced adversarial robustness compared to their teacher counterparts, suggesting that they may not deeply mimic the teacher’s learned representations and decision boundaries. However, this conclusion is based on experiments restricted to encoder-only models and classification-based tasks. Generation tasks and other transformer-based architectures (e.g., decoder-only and encoder-decoder models) were excluded because current state-of-the-art knowledge distillation methods for code language models primarily target encoder-only models and classification tasks. While recent work, such as SODA~\cite{chen2025smaller} investigates distillation for decoder-based models, a complete and reproducible implementation\footnote{https://shorturl.at/B8zW9} was not publicly available during the period of this study. Including such methods without full reproducibility would risk compromising experimental fairness and scientific rigor. To address this limitation, multiple widely used models, datasets, and knowledge distillation techniques were adopted under consistent experimental settings and evaluated using diverse adversarial attacks, complementary metrics, and the \textit{MetaCompress} framework. The consistent degradation in robustness observed across all evaluated settings strengthens confidence in these findings. Nevertheless, these conclusions may not fully generalize to generation tasks. Future research should extend this analysis to a broader range of model families and task types.

\section{Related Work}
\label{RW}

A substantial body of research on model compression has emerged in the fields of computer vision and natural language processing (NLP), aiming to develop large language models that are energy efficient, environmentally sustainable, and optimized for memory usage \cite{xu2023survey, zhu2024survey, xu2025resource, xu2021survey, gordon2020compressing, xu2021beyond, ye2019adversarial, du2021robustness, gourtani2024improving, zhu2022safety, sanh2019distilbert, sun2019patient, jiao2019tinybert, buciluǎ2006model, jiang2023lion, zhang2018structadmm, tang2019distilling, xu2020bert}. To this end, several techniques, including \textit{pruning} \cite{sanh2020movement}, \textit{quantization} \cite{zafrir2019q8bert}, and \textit{knowledge distillation} (KD) \cite{hinton2015distilling}, have been widely explored. More recently, these compression techniques have been extended to software analytics to improve the efficiency of language models of code. Wei et al. \cite{wei2023towards} evaluated quantized models on code generation tasks, examining accuracy, resource consumption, and environmental impact, and found that quantization can substantially improve model efficiency with minimal degradation in accuracy or robustness when applied under suitable conditions. Saad et al. \cite{saad2024alpine} introduced \textit{ALPINE}, a pruning method that is independent of programming language and effectively reduces the computational cost of code-language models. Aloisio et al. \cite{d2024compression} conducted an empirical study on how different compression strategies, such as pruning, quantization, and knowledge distillation, affect the performance and resource efficiency of the CodeBERT model across three software analytics tasks: code summarization, vulnerability detection, and code search. Afrin et al. \cite{afrin2025quantization} empirically examined the effects of quantizing large code models beyond functional correctness, focusing on qualitative aspects and static properties of the generated code. While pruning and quantization improve efficiency, their compression capacity is limited, as excessive compression often harms model performance. Hence, this study focuses on knowledge distillation, which enables extremely compact models without compromising significant predictive performance.

Shi et al. \cite{shi2022compressing} proposed \textit{Compressor}, a KD-based method evaluated on CodeBERT and GraphCodeBERT for two code understanding tasks: vulnerability prediction and clone detection. Experimental results show that \textit{Compressor} significantly reduces inference time and model size, down to 3 MB, while preserving model effectiveness. In a follow-up study, Shi et al. \cite{shi2024greening} introduced \textit{AVATAR}, another KD-based approach designed to mitigate energy consumption and carbon emissions, while also maintaining model effectiveness and reducing inference latency. Building on \textit{AVATAR}, \textit{MORPH} \cite{panichella2025metamorphic} was proposed as a KD method that combines metamorphic testing with many-objective optimization to enhance model robustness on metamorphic code, as well as accuracy, size, and efficiency. Experiments on CodeBERT and GraphCodeBERT showed that \textit{MORPH} achieved up to 73\% greater robustness and 38\% greater efficiency than \textit{AVATAR}, without sacrificing accuracy or increasing model size. However, in \textit{MORPH}, the metamorphic attack using a metamorphic sample technically differs from the adversarial attack using an adversarial example. Metamorphic attacks are performed only once on the metamorphic sample, whereas adversarial attacks are performed by repeatedly querying the model, modifying the original sample to assess the model's actual robustness. Our empirical evaluation also reveals that the student models obtained from \textit{MORPH} experience greater performance degradation under adversarial attacks than under metamorphic attacks, compared to the student models obtained from \textit{AVATAR}.

Recently, Liu et al. \cite{liu2026pioneer} proposed a KD-based approach, PIONEER, that enhances the robustness of compressed code models while preserving accuracy. Using a solver-constrained genetic algorithm and robustness evaluation with ALERT-generated examples, it achieved an 87.54\% robustness gain with only a 1.67\% drop in accuracy across CodeBERT, GraphCodeBERT, and CodeT5. Recently, Wang et al. \cite{wang2025empirical} empirically evaluated the effectiveness of logit-based and feature-based KD approaches for code understanding tasks. Their study showed that KD consistently outperforms fine-tuning on code understanding tasks, with feature-based methods and code-specific teachers yielding the best results.

Despite recent advances in applying knowledge distillation to software analytics, a key limitation remains in how distillation effectiveness is evaluated. Current methods predominantly utilize standard performance metrics, such as accuracy, precision, recall, F1-score, ROUGE, and BLEU-4, to compare student and teacher models. Although these metrics measure task-level correctness, they do not determine whether the student model mimics the teacher’s predictive behavior, confidence distribution, or response consistency across varying inputs. As a result, two models may exhibit comparable performance on benchmark datasets while differing substantially in their internal representations and robustness characteristics. This limitation is particularly critical in adversarial or safety-sensitive settings, where such behavioral discrepancies can lead to unreliable predictions. To date, no study has systematically examined behavioral fidelity between teacher and student models in this context. This gap highlights the need for a principled evaluation framework that goes beyond performance metrics to explicitly assess whether a student model faithfully mimics its teacher's behavioral characteristics, which we address through a metamorphic testing-based approach.

\section{Conclusion}
\label{ConClu}
This study investigated the extent to which smaller student models mimic the predictive behavior and internal representations of their larger teacher counterparts in knowledge distillation. Our findings showed that, although student models achieved performance comparable to that of their teachers, their performance deteriorated significantly under adversarial attacks. This degradation underscores their limited ability to deeply mimic the teacher models’ decision-making behavior and internal representations, a limitation that traditional accuracy-based evaluations fail to capture.

To address this limitation, we introduce \textit{MetaCompress}, a metamorphic testing framework that extends behavioral fidelity assessment beyond \textit{accuracy} by framing it as a classical software testing problem and applying metamorphic testing principles. Extensive experiments across diverse language models of code, software engineering tasks, and knowledge distillation techniques show that \textit{MetaCompress} effectively uncovers discrepancies in behavioral fidelity between teacher and student models that are often missed by traditional accuracy-based evaluations. Additionally, an ablation study demonstrates the robustness of \textit{MetaCompress} framework in detecting behavioral fidelity divergences across transformed inputs. As future work, we plan to integrate \textit{MetaCompress} into the knowledge distillation pipeline to enable feedback-driven optimization of the distillation process, aiming to reduce the knowledge and representational gaps in student models and ensure behavioral fidelity with their teacher counterparts.

\section*{Data Availability}
Our code and the corresponding dataset are publicly available to enhance further research\footnote{\url{https://doi.org/10.5281/zenodo.16127320}}.

\section*{Acknowledgments}
This research is supported in part by the Natural Sciences and Engineering Research Council of Canada (NSERC) Discovery Grants program, the Canada Foundation for Innovation's John R. Evans Leaders Fund (CFI-JELF), and by the industry-stream NSERC CREATE in Software Analytics Research (SOAR).

\section*{Declaration of generative AI and AI-assisted technologies in the manuscript preparation process}
During the preparation of this work, the author(s) used Grammarly \cite{Grammarly} and ChatGPT \cite{OpenAI2025} to find grammatical mistakes and improve sentence clarity/ presentation. After using these tools/services, the author(s) reviewed and edited the content as needed and take(s) full responsibility for the content of the publication.

\bibliographystyle{elsarticle-num}
\bibliography{cas-refs}

\end{document}